\documentclass[reprint,amsmath,amssymb,aps,showpacs]{revtex4-1}
\usepackage{graphicx}
\usepackage{dcolumn}
\usepackage{bm}
\usepackage{float}
\usepackage{epstopdf}
\usepackage{amsmath}
\usepackage{amsfonts}
\usepackage{amssymb}
\date{today}
\newcommand{\be}{\begin{equation}}
\newcommand{\ee}{\end{equation}}
\newcommand{\bea}{\begin{eqnarray}}  
\newcommand{\eea}{\end{eqnarray}}

\newcommand{\p}{\partial}
\newcommand{\s}{\sigma}

\newcommand{\up}{\uparrow}
\newcommand{\down}{\downarrow}
\newcommand{\la}{\langle}
\newcommand{\ra}{\rangle}
\newcommand{\rd}{\mbox{d}}
\newcommand{\ri}{\mbox{i}}

\newcommand{\nn}{\nonumber}
\newcommand{\bp}{\bar{\partial}}
\newcommand{\bz}{\bar{z}}
\newcommand{\bw}{\bar{w}}

\newcommand{\vare}{\varepsilon}

\newcommand{\vvphi}{\mbox{\boldmath $\varphi$}}

\newcommand{\vnab}{\mbox{\boldmath $\nabla$}}

\newcommand{\htau}{\hat{\tau}}

\newcommand{\bcr}{\begin{array}{clcr}}
\newcommand{\ecr}{\end{array}}

\newcommand{\fU}{\cal{U}}

\newcommand{\sx}{\sin x}
\newcommand{\sy}{\sin y}
\newcommand{\cx}{\cos x}
\newcommand{\cy}{\cos y}

\newcommand{\frm}{\mathfrak{m}}
\input{epsf}

\begin{document}

\title{Ground state phases and quantum criticalities of one-dimensional Peierls model with
spin-dependent sign-alternating potentials}
\author{G.A.Japaridze$^{1,2}$}
\author{A.A.Nersesyan$^{1,2,3}$}
\affiliation{\small{\sl $^1$Ilia State University, 0162, Tbilisi, Georgia}}
\affiliation{\small {\sl $^2$Andronikashvili Institute of Physics, 0177, Tbilisi, Georgia}}
\affiliation{\small {\sl $^3$The Abdus Salam International Centre for Theoretical 
Physics, 34151, Trieste, Italy}}
\date{\today}

\begin{abstract}

We consider 
a one-dimensional 
commensurate Peierls insulator in the presence of spin-dependent sign-alternating
potentials. 
In a continuum description, the latter supply the fermions with spin-dependent
"relativistic" masses $m_{\up,\down}$.
The ground-state phase diagram 
describes three gapped phases: the CDW and SDW-like band insulator phases sandwiched by
a "mixed" phase in which the CDW and SDW superstructures
coexist with a nonzero spontaneous dimerization (SD). 
The critical lines separating the massive phases  belong to the Ising universality 
class. 
The 
Ising criticality is accompanied by the Kohn anomaly in the renormalized phonon spectrum.  We derive a Ginzburg criterion which specifies a narrow region around the critical point where quantum fluctuations play a dominant role, rendering the 
adiabatic (or mean-field) approximation inapplicable.

A full account of quantum effects is achieved in the anti-adiabatic limit where the effective
low-energy theory represents a massive version the N=4 Gross-Neveu model.
Using Abelian bosonization we
demonstrate that the description of the SD phase, including its critical boundaries, 
is well approximated by a sum of two 
effective double-frequency sine-Gordon (DSG) models subject to self-consistency conditions that couple the charge and spin sectors. Using the well-known critical properties of the DSG model we 
obtain the singular parts of the dimerization order parameter and staggered
charge and spin susceptibilities near the Ising critical lines.
We show that, in the anti-adiabatic limit, on the line $m_{\down} = 0$ 
there exists of a Berezinskii-Kosterlitz-Thouless critical point separating a Luttinger-liquid gapless phase from the spontaneously dimerized one.
We also discuss topological excitations of the model carrying fractional charge and spin.

\end{abstract}


\maketitle

\section{Introduction}

The effects of structural distortions or external 
symmetry breaking fields
on the properties of strongly correlated electron systems have long been the
subject of thorough investigations in condensed matter physics.
A prototype model
to study these effects is the one-dimensional (1D) ionic Hubbard model, which is a repulsive Hubbard chain
at 1/2 filling with a 
staggered scalar potential.
This model was originally proposed \cite{nagaosa} for the description of 
organic mixed-stack charge-transfer crystals with alternating donor and acceptor
molecules;  later it has been discussed in the context of ferroelectricity in
transition metal oxides \cite{egami}. 
In view of the
general interest in quantum phase transitions between gapped  phases
with qualitatively different symmetry properties, the main theoretical question here
concerns the nature of the crossover between the Mott Insulator (MI) 
and Band Insulator (BI) phases, taking place when 
at a fixed amplitude of the staggered potential
the on-site Coulomb repulsion $U$
It has been demonstrated on the basis of field-theoretical arguments 
\cite{FGN1,FGN2} that there is no direct transition between the MI and BI ground state phases.
In fact, the MI-BI crossover 
is realized as a sequence of two continuous quantum phase
transitions separated by 
an intermediate, fully gapped, long-range ordered phase characterized by a
spontaneous dimerization (SD) (this phase is also called bond-order wave). The MI-SD transition ($U = U_{c2}$) is associated with opening of
a spin gap and belongs to the Berezinskii-Kosterlitz-Thouless (BKT) universality class, whereas
the SD-BI criticality ($U_{c1}$) occurs in the charge part of the spectrum and is of the 
quantum Ising type.
In spite of some 
controversy in subsequent results that followed
this prediction \cite{FGN1,FGN2}, the two-transition scenario 
is now well documented  in both analytical and numerical works  \cite{manmana,zhang,dyonis}.
The renewal of interest in the ionic Hubbard model is caused by its recent realization on optical lattices of ultracold fermionic
atoms \cite{tarruell, exp1, messer} and the attempts to detect the predicted spontaneously dimerized phase
sandwiched between the MI and BI states \cite{orignac}. A SD 
phase has also been
shown to exist in the extended 1/2-filled Hubbard model which includes a nearest-neighbor
repulsion between the electrons (the UV model)\cite{nakamura,furu}.

In the 1D ionic Hubbard model defined on a rigid lattice, the region where both transitions occur 
is determined by the single condition 
that the commensurability gap of the spectrum of the MI 
and the one-particle gap of the BI 
are of the same order. As a result,
the SD phase occupies a very narrow region in the phase diagram, the fact that complicates
its experimental detection. However, it has been pointed out  \cite{FGN1,{zhang}} 
that inclusion of the electron-phonon interaction 
moves the Ising and BKT critical boundaries apart, thus removing ambiguities about the nature of the MI-BI
crossover. Moreover, if the electron-phonon coupling is strong enough, the MI  phase
ceases to exist, and the model displays only the SD and BI phases. Thus the ionic  Peierls-Hubbard
model is more
appropriate for a detailed study of quantum phase transitions in 1D systems controlled by
a combined effect of electron-electron correlations, sign-alternating potential and electron-phonon coupling.

Spin degrees of freedom play essential role in the formation of strongly correlated phases of electrons in
one dimension. Therefore, apart from the ionic staggered potential that acts on the electron 
charge density, the presence of a staggered magnetic field is also expected to affect the ground state
phase diagram of the system. It should be pointed out that including a staggered magnetic field into
consideration is not only of a theoretical interest. Experimental realizations of effective
internal magnetic fields alternating over an atomic scale in certain quasi-1D compounds
have already been reported in the literature \cite{dender,aff-oshi,ess-tsv,zheludev}. We believe that
artificial manufacturing and controllable manipulation of spin-dependent potentials 
for ultracold atoms 
on optical lattices (for instance, in systems with mass-imbalanced atomic species)
is a feasible task which will be performed before long. 
Thus it makes sense to study a more
representative model of correlated fermions which incorporates staggered potentials depending on the spin projection
of the particles.

As a first step in this direction, in this paper we address the role of a spin-dependent
sign-alternating potential in the Peierls model \cite{Peierls} which ignores the direct on-site Coulomb repulsion 
between the fermions ($U=0$) but accounts for those correlations between the particles which are
mediated by electron-phonon coupling. 
In the context of the Hubbard-Peierls model, such approximation can be justified 
by the assumption that the phonon-mediated attraction between the electons
is stronger than the Coulomb repulsion $U$ (see section \ref{anti-adiabat}).
The model is described by the Hamiltonian
\bea
H = &-& \sum_{n,\s} \Big[ t_0 + \frac{1}{2} (-1)^n \hat{\Delta}  \Big] \left( 
c^{\dagger}_{n\s} c_{n+1, \s}
+ h.c. \right)\nn\\
&+& \sum_{n,\s} m_{\s} (-1)^n c^{\dagger}_{n\s} c_{n\s} 
+ H_{\rm ph}, \label{ion_pier_spinless}\\
H_{\rm ph} &=& \frac{1}{2g^2 _0} \int \rd x~ \Big[ \frac{1}{\omega^2 _0} 
\left( \frac{\p \hat{\Delta}}{\p t} \right)^2 + \hat{\Delta}^2 \Big].
\label{ph-ham}
\eea
Here $c^{\dagger}_{n\s}(c_{n\s})$ is the creation (annihilation) operator for an electron
on site $n$ with spin projection $\s=\up,\down$, $t_0$ is the  hopping matrix element, and
$\hat{\Delta}$ is a real, dispersionless 
quantum displacement field describing phonon modes
with wave vectors close to $q = \pi/a_0$ ($a_0$ being the lattice spacing).
The tight-binding band of the electrons is assumed to be exactly 1/2-filled.
The electron-phonon coupling is derived from the first-order modulation of the nearest-neighbor hopping
amplitudes due to longitudinal lattice vibrations -- the 
Su-Schrieffer-Heeger (SSH) model \cite{SSH, ssh2}.
$g^2 _0$ is the electron-phonon coupling constant.
The amplitudes of spin-dependent one-particle potentials $m_{\s}$ 
determine the scalar potential $m_+ = (m_{\up} + m_{\down})/2$ which 
supports a site-diagonal CDW ordering, and the staggered magnetic field $m_- = (m_{\up} - m_{\down})/2$ which tends to induce a Neel (SDW) alignment of fermionic spins.
In the SSH model, the phonon field  $\hat{\Delta}$ couples to the electron dimerization operator
\be
D_{n,n+1}  = \frac{1}{2}(-1)^n \sum_{\s} \left( c^{\dagger}_{n\s}  c_{n+1, \s} + h.c. \right)
~~~
\label{D}
\ee
whose expectation value is the SD order parameter.

At $m_{\pm} = 0$ and arbitrary nonzero value of
 $g_0$,
the SSH model (\ref{ion_pier_spinless}) has a doubly degenerate SD ground state
and 
adequately describes
structural and electronic properties of the trans-polyacetylene (CH)$_x$.
The charge-spin separated 
nature of topological soliton 
excitations is the most celebrated feature of this model \cite{SSH,ssh2,braz1}. The case of a scalar staggered potential
$m_+ \neq 0, m_- = 0$ corresponds to a 
Peierls insulator with a broken charge conjugation symmetry \cite{ns}; it is 
relevant to the polymer cis-(CH)$_x$
\cite{bk, bkm}.

The dimerization field (\ref{D}) and the staggered electron charge or spin densities,
$\rho_{s} (n) = (-1)^n \sum_{\s}c^{\dagger}_{n\s} c_{n\s}$ or
$\s^z _s (n) = (-1)^n \sum_{\s}\s c^{\dagger}_{n\s} c_{n\s}$, 
have different parity propertis. $D_{n,n+1}$ is invariant under link parity transformation 
$
P_L ~(n \to 1-n)
$
but changes its sign under site parity 
$
P_S ~(n \to - n),
$
whereas for $\rho_{s} (n)$ and $\s^z _s (n)$
the situation is just the opposite. 
Therefore, 
starting 
from the spontaneously dimerized phase of a Peierls Insulator (PI) 
at $m_{\pm} = 0$ and increasing the
staggered amplitudes, one could trace a crossover to the BI regime 
realized as a quantum criticality. 
Another argument in favor of this scenario is based
on the observation that the PI massive phase is an example of a topological insulator \cite{shen}, whereas
the BI described by the Hamiltonian (\ref{ion_pier_spinless}) at $\Delta = 0$ is not.

We would like to stress here that, as opposed to earlier papers \cite{furu}, \cite{essler-2006}
where the electron bond dimerization competing with the site-diagonal CDW potential was assumed to be {explicit}, everywhere in the present paper we will be dealing with
\emph{spontaneous} dimerization. 

In the alternative, Holstein model, which is more appropriate for molecular crystals, the displacement field $\hat{\Delta}$ of the Einstein phonons couples
to the staggered part of the electron charge density, 
$\hat{\Delta} \sum_{n\s} (-1)^n c^{\dagger}_{n\s} c_{n\s}$.
In such model 
the role of the staggered potentials is more trivial than in the SSH model 
(\ref{ion_pier_spinless}) because the PI phase itself and the perturbing potentials $m_{\s}$
are all $P_S$-symmetric. In this case, the staggered potentials simply
lift the $\mathbb{Z}_2$-degeneracy of the PI state. No phase transitions are expected
in this case. We thereby will not consider the Holstein model 
in what follows.

Throughout this paper we will be concerned with the weak-coupling limit in which
all parameters of the microscopic model (\ref{ion_pier_spinless}) with the dimension of energy
are much less than the bandwidth $4t_0$:
\be
|\Delta|, |m_{\s}| \ll W, ~~~~\lambda_0 = 2g^2 _0 /\pi v_F \ll 1
\label{infrared}
\ee
Here $v_F = 2t_0 a_0$ is the Fermi velocity, $W \sim v_F / a_0$ is the high-energy cutoff of the electron spectrum, and $\lambda_0$ is the dimensionless electron-phonon coupling constant. In this case only electron states close to
the right and left Fermi points are important. Being interested in
the low-energy properties of the system, in the fermionic part of the Hamiltonian, 
Eq.(\ref{ion_pier_spinless}),
one can pass to the continuum limit
\bea
 {\cal H}_f (x)  = 
\sum_{\s}\psi^{\dagger}_{\s} (x)\left( - \ri v_F \hat{\tau}_3 \p_x 
+ \Delta \htau_2 + m_{\s} \htau_1
\right)
\psi_{\s} (x)~~~~
\label{ham-cont-lim}
\eea
Here
\[
\psi_{\s} (x)= \left(
\begin{array}{clcr}
R_{\s} (x)\\
L_{\s} (x)
\end{array}
\right)
\]
is a 2-spinor whose components are right and left chiral fermionic fields. 
The Pauli matrices
$\htau_i~(i=1,2,3)$ act in the two-dimensional Dirac-Nambu space. 
The quantum order parameter
field ${\Delta}$ couples to the electron dimerization operator
\be
D (x) = \sum_{\s}\psi^{\dagger}_{\s} (x) \hat{\tau}_2 \psi_{\s} (x)\label{D-cont}
\ee
which is the continuum version of (\ref{D}).

At 
$\Delta = 0$ ${\cal H}_f$ in (\ref{ham-cont-lim}) is a sum of two
Dirac models with spin dependent masses $m_{\s}$
accounting for external backscattering potentials. This is a BI model.
At $m_{\up,\down} = 0$ the Hamiltonian (\ref{ham-cont-lim}),  with the phonon contribution
(\ref{ph-ham}) included, represents the continuum version of the 
commensurate Peierls model,
introduced by Takayama, Lin-Liu and Maki \cite{tll} 
(a field-theoretical description of an incommensurate Peierls model was developed earlier
in Ref.\cite{BD})
and subsequently analyzed in great detail
by Fradkin and Hirsch \cite{fh1}.
A general feature of this model is 
dynamical generation of an exponentially small
spectral gap and two-fold degeneracy of the
spontaneously dimerized ground state.
The study of the ground-state phase diagram of the model (\ref{ion_pier_spinless}),
(\ref{ph-ham}) which results from the competition between the external potentials
$m_{\s}$ and electron-phonon interaction $g^2 _0$ is the main goal of the present paper.

The paper is organized as follows. In sec.2 the model is considered within a semi-classical,
adiabatic approximation. We show that, as long as $m_{\up}, m_{\down} \neq 0$,
there exists a threshold for electron-phonon coupling 
at which an Ising transition to the spontaneously dimerized phase 
takes place. Taking into account quantum fluctuations within Random Phase Approximation
we demonstrate the existence of a Kohn anomaly in the renormalized phonon spectrum
at the critical point. We also derive a Ginzburg criterion which determines the range of
applicability of the adiabatic approximation close to the transition point. In sections 3, 4 we 
discuss the effective low-energy model emerging in the anti-adiabatic limit, which is the
case of high-frequency quantum phonons. We bosonize this model in terms of the scalar fields 
$\Phi_c$ and $\Phi_s$ describing collective charge and spin excitations. We identify
stable minima of the potential $U(\Phi_c, \Phi_s)$, which incorporates all perturbations to
the Gaussian models of the charge and spin sectors, and trace transformations of these minima
as the parameters of the model are varied. Based on this analysis we provide a
qualitative ground-state phase diagram of the model (Fig.\ref{antiadiabatic-phase-dia}).
We propose
a well justified approximate scheme in which our bosonized model
is reduced to a sum of double-frequency (DSG) models in the charge and spin sectors,
coupled by self-consistency conditions. 
Using this picture 
and the known results on the DSG model we estimate the
singular parts of the dimerization order parameter and the staggered charge and spin
susceptibilities close to Ising criticalities. We also classify topological excitations of the model, the fractional quantum numbers they carry, and their evolution under the change of
the parameters of the model. In sec.5 we consider a special case when a staggered potential
is applied to one fermionic spin component only ($m_{\up} \neq 0$, $m_{\down} = 0$).
By integrating out the massive degrees of freedom we derive the effective bosonized action
for the spin-$\down$ fermions and discuss its properties.

The paper contains three appendices. In Appendix A we provide necessary details
of Abelian bosonization used in the main text. 
In Appendix B we analyze stable vacua of the potential of the effective bosonized
model appearing in the anti-adiabatic limit (sections 3,4). Appendix C contains
technical details relating to the derivation of the effective action in sec.5.

\section{Adiabatic limit}
\subsection{Mean-field theory}\label{mft}

In this section we adopt the
adiabatic approximation in which
the $\pi$-phonons are treated semiclassically. In this approach,
equivalent to
a mean-field theory, 
quantum fluctuations of the order parameter field ${\Delta}$ are neglected, so $\Delta$
becomes
a classical 
variational parameter whose equilibrium value is found by minimizing the ground state energy of the electron-phononsystem
\be
E_0 (\Delta) = - \sum_{k,\s} \sqrt{k^2 v^2 _F + m_{\s}^2 + \Delta^2} + \frac{L \Delta^2}{2g^2 _0}
\ee
Simple calculations lead to 
\bea
&&
{\cal E} (\Delta) \equiv L^{-1}
\left[ E (\Delta) -E (0)\right]\nn\\
&& = 
\frac{\Delta^2}{2g_0 ^2}
 - \sum_{\s}
\frac{{\Delta}^2+ m^2 _{\s}}{2\pi v_F}
\left( \ln \frac{W}{\sqrt{{\Delta}^2+ m^2 _{\s}}}+ \frac{1}{2}\right)
\label{tot-ene-mf}
\eea
Minimizing (\ref{tot-ene-mf}) 
we obtain 
\bea
\Delta \left( 1 - \frac{\lambda_0}{2} \sum_{\s}
\ln \frac{2W}{\sqrt{{\Delta}^2+ m^2 _{\s}}} \right) = 0
\label{EQUA}
\eea
Re-expressing $\lambda_0$ in terms of the Peierls gap $\Delta_0$, corresponding to the unperturbed case ($m_{\up} = m_{\down} = 0$),
\[
{1}/{\lambda_0} = \ln ({2W}/{|\Delta_0|})
\]
we find that 
in the region $|m_{\up} m_{\down}| < \Delta_0 ^2$
Eq.(\ref{EQUA}) leads to a doubly degenerate
nonzero  equilibrium solution $\pm \Delta$ satisfying
\be
\Delta^2 = - \frac{m_{\up}^2 + m^2 _{\down}}{2}
+ \sqrt{\frac{(m_{\up}^2 - m^2 _{\down})^2}{4} + \Delta_0 ^4},
\label{new-Delta}
\ee
whereas at $|m_{\up} m_{\down}| > \Delta_0 ^2$ only a trivial solution $\Delta = 0$
remains.
\begin{figure}[hbbp]
\centering
\includegraphics[width=2.3in]
{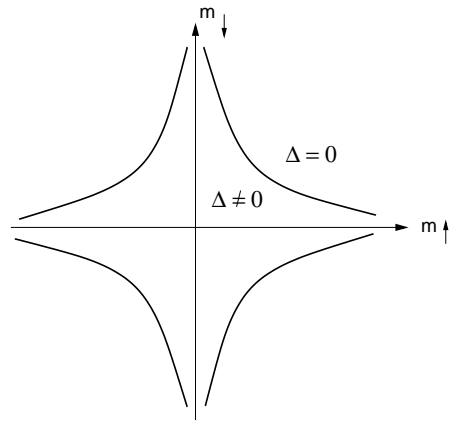}
\caption{\footnotesize Mean-field ground-state phase diagram.
The critical lines are determined by the equation
$|m_{\up}m_{\down}| = \Delta^2 _0$.}
\label{mf-phase}
\end{figure}
\noindent
The curves
$
|m_{\up} m_{\down}| = \Delta_0 ^2\label{crit-lines}
$
represent critical lines separating the ordered, spontaneously dimerized phase
from the disordered phase 
representing a non-dimerized band insulator (see Fig.\ref{mf-phase}). 
In the vicinity of these lines
we can set
\be
|m_{\up} m_{\down}| = \Delta_0 ^2 (1 + \tau), ~~~
|\tau| \ll 1 \label{vicinity}
\ee
Then
the ground state energy 
takes the form of a $\mathbb{Z}_2$-symmetric
Landau  expansion for a \emph{classical} Ising model
\bea
{\cal E}(\Delta) &=& \frac{1}{\pi v_F}
\left( \frac{1}{2} \tau \Delta^2 + \frac{\Delta^4}{4 m^2 _*} + \cdots
\right), \label{landau-rexpan}\\
\frac{1}{m^2 _*} &=& \frac{1}{2} \sum_{\s} \frac{1}{m^2 _{\s}}
\label{inverse-m2}
\eea

In special cases $m_{\up} = \pm m_{\down} \equiv m$ corresponding to
pure CDW or SDW external potentials the nontrivial solution reduces to
$
\Delta = \pm \sqrt{\Delta^2 _0 - m^2}
$
and exists if  $|m| < |\Delta_0|$. The conditions for the onset of SD
in the ionic Peierls model ($m_{\up} = m_{\down} \neq 0$)
have been discussed earlier in Refs. \cite{bk, bkm}.

In the ordered phase ($\Delta \neq 0$), 
the ground-state average values of the
spin-symmetric 
and antisymmetric 
parts of the electron dimerization are 
\bea
D_{+}&=& \sum_{\s} \la  D^{\s}_{n,n+1} \ra \simeq \mp \frac{2 \Delta }{g^2 _0}, 
\label{bcdw}\\
D_{-}&=& \sum_{\s} \s \la  D^{\s}_{n,n+1} \ra \simeq \pm \frac{\Delta}{\pi v_F}
\ln \frac{m^2 _{\up} + \Delta^2}{m^2 _{\down} + \Delta^2}
\label{bsdw}
\eea
The average staggered parts of the electron density for a given $\s$ are given by
\bea
\rho_{\rm stag}^{\s} &=& \frac{1}{N}\sum_n (-1)^n \la c^{\dagger}_{n\s} c_{n\s} \ra \nn\\
&=& - \frac{m_{\s}}{\pi v_F} \ln \frac{W}{\sqrt{m^2 _{\s} + \Delta ^2}}
\label{stag-dens}
\eea

Thus, in the presence of spin-anistropic staggered potential,
the system displays two gapped phases: a non-dimerized  band-insulator phase with site-diagonal
charge and spin density waves 
and a mixed phase with a doubly degenerate ground state, in which a nonzero spontaneous
dimerization coexists with the explicit CDW and SDW superstructures.
As long as both $m_{\up}$ and $m_{\down}$ are nonzero, there exists a threshold for
electron-phonon coupling, 
$\lambda_{c} =  \ln (2W /\sqrt{|m_{\up} m_{\down}|})$,
at which an Ising transition to the spontaneously dimerized
phase ($\lambda_0 > \lambda_{c}$) takes place.
The case when one of the alternating amplitudes is zero (i.e. one fermionic spin component
being massless) 
will be considered separately in Sec. \ref{1-comp}.

\subsection{Phonon dynamics in RPA. Kohn anomaly and Ginzburg criterion}

In the previous subsection we adopted the conventional mean-field approach to derive the Peierls dimerization gap in the one-particle spectrum. A consistent extention of this 
method to the two-particle level which allows one to include quantum fluctuations above the mean-field solution is the Random Phase Approximation (RPA) applied to the 
electron dynamical susceptibility and the phonon Green's function. In doing so we will follow the seminal
works by Lee, Rice and Anderson \cite{LRA} and by Brazovskii and Dzyaloshinskii \cite{BD}. 
We will show that close to
the mean-field Ising transition the phonon spectrum exhibits a Kohn anomaly -- softening
of the phonon
optical gap at $q = \pi$. Exactly at the critical point the $\pi$-phonons become gapless.
We will derive a Ginzburg criterion which establishes the range of applicability
of the mean-field (adiabatic) approximation in the vicinity of the transition.

Let us separate the mean-field value of the quantum field $\hat{\Delta}$ and its fluctuation
(normal ordered part),
$
\hat{\Delta}(x,t) = \Delta + \hat{\delta} (x,t)$, $\Delta \equiv \la  \hat{\Delta}(x,t)\ra.
$
The expectation value $\Delta$ enters the mean-field fermionic Hamiltonian
\bea
H_f &=& \sum_{k\s} \psi^{\dagger}_{k\s} \hat{\cal H}_{k\s} \psi_{k\s},
~~\psi_{k\s} = \left(
\begin{array}{clcr}
R_{k\s}\\
L_{k\s}
\end{array}
\right)\nn\\
&& \hat{\cal H}_{k\s} = kv_F \htau_3 + \Delta \htau_2 + m_{\s} \htau_3
\label{ham-k}
\eea
The remaining, normal ordered  part of the electron-phonon interaction 
\be
:H_{\rm e-ph} : ~= \int \rd x~ \hat{\delta} (x) :\hat{D}(x):
\label{e-ph1}
\ee
accounts for quantum fluctuations
of the order parameter field. 
Introduce the Matsubara Green's function (GF) of the optical phonon
\be
{\cal G} (x, \tau) = \la T_{\tau} :\hat{\Delta} (x,\tau): :\hat{\Delta} (0,0): \ra
\label{G-phon-def}
\ee
where $T_{\tau}$ is the imaginary-time ordering operator and averaging goes over the ground state of the system. The Fourier transform of this function,
\be
{\cal G} (q,\omega) =  \int \rd x  \int_{-\infty} ^{\infty} \rd \tau~
{\cal G} (x, \tau) ~e^{-iqx + i \omega \tau}\label{ph-GF-four}
\ee
satisfies the Dyson equation \cite{AGD}
\be
{\cal G}^{-1} (q, \omega) = {\cal G}^{-1} _0 (\omega) - X(q, \omega)
\label{dyson-exact}
\ee
where $X(q, \omega)$ is the exact dynamical dimerization susceptibility
(or polarization operator) defined as the Fourier transform of the two-particle GF
\bea
 X(x,\tau) = 
\la :\hat{D} (x,\tau): :D(0,0):  \ra
\label{X-def}
\eea
The momentum $q$ is measured from the value $2k_F = \pi/a_0$.
In Eq.(\ref{dyson-exact}) 
$
{\cal G} _0 (\omega) = g^2 _0 \omega^2 _0 /(\omega^2 + \omega^2_0)$
is the GF of the bare dispersionless optical phonon. Its spectrum gets renormalized
due to quantum polarization effects of the electron subsystem incorporated in
$X(q,\omega)$.
In the RPA we adopt here, $X(q,\omega)$ is replaced by its value for free massive
fermions described by the Hamiltonian (\ref{ham-k}):
\bea
X^{(0)}(q,\omega) &=& -  \sum_{\s}\int \frac{\rd k}{2\pi} \int \frac{\rd \vare}{2\pi}
\nn\\
&\times&{\rm Tr}~ 
\{ \hat{G}_{\s}(k+q, \vare + \omega) \htau_2 \hat{G}_{\s} (k,\vare) \htau_2 \},~~
\label{X2}
\eea
where 
\bea
\hat{G}_{\s} (k,\vare)
= - \frac{\ri \vare + kv_F \htau_3 + \Delta \htau_2 + m_{\s} \htau_1}
{\vare^2 + k^2 v^2 _F + \Delta^2 + m^2 _{\s}},
\label{f-GF}
\eea
is the single-fermion 2$\times$2 GF matrix.

Staying within the adiabatic approximation, we will be interested in
the phonon dynamics in the low-energy limit:
$|\omega| \ll |\Delta|$, $|m_{\s}|$.
A straightforward calculation shows that in this region
\bea
X^{(0)}(q, \omega) &=& \frac{1}{\pi v_F}\sum_{\s}
\Big[ \ln \frac{2W}{\sqrt{\Delta^2 + m^2 _{\s}}} -1\nn\\
 &&~~~~~~- ~ 
\frac{\omega^2 + q^2 v^2 _F - 4 m^2 _{\s}}{4 (\Delta^2 + m^2 _{\s})}
\Big]
~~~~\label{chi-ret-qomega-mats}
\eea
Then, in the leading order, taking into account the mean-field equation
(\ref{EQUA}), we obtain 
\bea
{\cal G}_{\rm RPA} (q, \omega) =
\frac{g^2 _0 \omega^2 _0}{\omega^2 + \Omega^2 + q^2 u^2}
\label{G-RPA}
\eea
where
\bea
\Omega = 
\Omega_0 \zeta, 
~~~~u  = {v_F \Omega}/{|\Delta|}, \label{Omega-u}
\eea
Here $\Omega_0 = \sqrt{\lambda_0} \omega_0$ is the optical gap
in the standard ($m_{\s} = 0$)
adiabatic Peierls model \cite{LRA, BD}.
Eq.(\ref{G-RPA}) describes an optical phonon with
the renormalized spectral gap $\Omega$ 
and group velocity $u$.
The parameter 
\[
\zeta = \left[ \frac{1}{2} \sum_{\s} \frac{\Delta^2}{\Delta^2 + m^2 _{\s}}\right]^{1/2}
\]
accounts for the interplay between the order parameter $\Delta$ and staggered
amplitudes $m_{\s}$ and
varies within the interval $0<\zeta<1$. 
At $m_{\s} = 0$ $\zeta = 1$, and formulas (\ref{G-RPA}),(\ref{Omega-u}) reproduce
the well-known results for the fluctuation spectrum 
of commensurate
Peierls systems \cite{LRA,BD}. At $\Delta \ll |m_{\s}|$, the parameter 
$\zeta \sim \Delta$. In the limit $\Delta 
\to 0$ the velocity $u$ remains finite; however  the gap $\Omega$ vanishes linearly with 
$\Delta$. This is a manifestation of the \emph{Kohn anomaly} in the phonon spectrum
signaling the onset of criticality.

Staying in the spontaneously dimerized phase, let us inspect the vicinity of the transition point more closely. In this region, using parametrization (\ref{vicinity})
with $\tau < 0$ and Eq.(\ref{new-Delta}), we find that $\Delta = m_* \sqrt{|\tau|}$.
So close to the critical point $\zeta = \sqrt{|\tau|}$ the renormalized optical
gap vanishes as 
$
\Omega = \Omega_0 \sqrt{|\tau|}.
$
The consistency requirement for the adiabatic approximation assumes relative smallness
of quantum fluctuations, i.e. $\la \hat{\delta}^2 (0,0) \ra \ll \Delta^2$. Using
(\ref{G-RPA}) we obtain
\bea
\frac{\la \hat{\delta}^2 (0,0)\ra}{\Delta^2} 
&=& \frac{1}{\Delta^2}\int \frac{\rd q}{2\pi} \int \frac{\rd \omega}{2\pi}~ {\cal G}_{RPA} (q,\omega)\nn\\
&=&  \frac{\sqrt{\lambda_0} \omega_0}{2\Delta \zeta} \ln \left(\frac{2W}{\sqrt{\lambda_0}
\omega_0 \zeta}\right)
\label{ave-fluct}
\eea
Deep in the ordered phase, i.e. at $\Delta_0 \gg |m_{\up} m_{\down}|$,
when $\Delta \sim \Delta_0$, the parameter $\zeta \sim 1$ and (\ref{ave-fluct})
reduces to the standard estimate of the ratio $\la \delta^2 \ra / \Delta^2$ for a conventional
PI. In this case the condition of the applicability of the adiabatic approximation
reads $(\Omega_0 / \Delta_0) \ln (2W/\Omega_0) \ll 1$.
The situation changes in the vicinity of the phase transition. At $\tau \to 0$
\bea
\frac{\la \hat{\delta}^2 \ra}{\Delta^2} \sim \frac{\Omega_0}{m_*} \frac{\ln (1/|\tau|)}
{|\tau|}
\label{ratio-crit}
\eea
Since $|\tau| \ll 1$, the ratio (\ref{ratio-crit}) can be small
only in the region
\be
1 \gg |\tau| \gg {\rm Gi} \label{Gi1}
\ee
where (with the logarithmic accuracy)
\be
{\rm Gi} \sim \frac{\Omega_0}{m_*} \ln \frac{m_*}{\Omega_0}
\label{Gi2}
\ee
is the Ginzburg parameter. The inequality (\ref{Gi1}) only makes sense if ${\rm Gi}\ll 1$,
i.e. $\Omega_0 \ll m_*$. If the latter condition is satisfied,
(\ref{Gi1}) represents the \emph{Ginzburg criterion} for our problem which
specifies the range of applicability of the semiclassical adiabatic (or mean-field) approximation near the critical point (see e.g.\cite{cardy}).
However, inside the narrow region
$
|\tau| \ll {\rm Gi} \ll 1 
$
the order parameter field cannot be be treated classically, and quantum fluctuations
dominate the dynamics of the system.

Thus one arrives at the important conclusion:
while away from criticality the adiabatic approximation is valid as long as
$\Omega _0 \ll |\Delta|, |m_{\s}|$, the immediate vicinity of the critical
point ($|\tau| \ll {\rm Gi}$) 
remains beyond the reach of
the above semiclassical approach.
An adequate route to correctly
account for strong quantum fluctuations of the order parameter is
the anti-adiabatic approximation which is discussed in detail in the remainder
of this paper. In particular, we will show that the classical Ising transition as it is seen 
from the "adiabatic distance" (\ref{Gi1}), transforms to a quantum criticality
belonging to the \emph{quantum Ising} universality class when quantum fluctuations
become dominant.

\subsection{Topological solitons}

We conclude this section by commenting on the fermion quantum numbers of 
solitons of the
field $\Delta(x)$, that is the coordinate dependent solutions that
interpolate between the two degenerate vacua $\pm {\Delta}_*$
of the potential (\ref{tot-ene-mf}) in the spontaneously dimerized phase:
\[
\lim_{x \to \infty} \Delta(x) = \pm {\Delta}_*, ~~~~\lim_{x \to - \infty} \Delta(x) = 
\mp {\Delta}_*
\]
At $m_{\up} = m_{\down} = 0$ the Hamiltonian
(\ref{ham-cont-lim}) anticommutes with $\htau_1$ and, hence, displays a conjugation
symmetry
$\psi \to \htau_1 \psi$ which takes positive-energy solutions of the Dirac equation into negative-energy
ones.
As a consequence, in the presence
of solitonic background, 
there exist normalizable, self-conjugate
zero-energy solutions of the Dirac equation carrying fractional quantized fermion numbers 
$q = \pm 1/2$. This was established almost simultaneously in quantum field theory \cite{jr}
and condensed matter physics \cite{SSH, ssh2,braz1}.
The mass term of the Dirac Hamiltonian breaks the charge conjugation symmetry. As a result,
for spinless fermions, the fermion
number $q$ is no longer half-integer but depends on the parameters of the Hamiltonian \cite{gw, ns}:
\[
q =  1/2 - (1/\pi) \tan^{-1} ({m}/{\Delta}_*),
\]
$m$ being the Dirac mass.
Since the Hamiltonian is additive in spin indices,
one
can use the above result 
for each fermionic spin component because
for both of them the topological background field $\Delta(x)$
is the same. Then one obtains
the charge and spin
quantum numbers for the topological solitons of the model (\ref{ion_pier_spinless}):
\bea
Q 
&=& 1 - \frac{1}{\pi}
\left[   \tan^{-1} \left(\frac{m_{\up}}{\Delta_*}\right) 
+  \tan^{-1} \left(\frac{m_{\down}}{\Delta_*}\right) \right] 
\label{adia-Q*}\\
S^z 
&=&  - \frac{1}{2\pi}\left[   \tan^{-1} \left(\frac{m_{\up}}{\Delta_*}\right) 
-  \tan^{-1} \left(\frac{m_{\down}}{\Delta_*}\right) \right]
\label{adia-S*}
\eea

For arbitrary nonzero 
values of the staggered amplitudes, such that $|m_{\up}|
\neq |m_{\down}|$,
topological excitations in our model carry both charge and spin which
continuously depend on the ratios $m_{\up, \down}/{\Delta}_*$. The cases 
$m_{\up} = \pm m_{\down}$
are special. The former case describes an ionic spin-symmetric PI in which
the staggered potential affects only the electron charge density. In such system
the solitons carry a nonzero fractional charge, $Q\neq 0$, but are spinless, $S^z = 0$.
The latter case corresponds to a staggered magnetic field affecting only the spin density
of the electrons. Accordingly, the solitons carry a nonzero spin, $S^z\neq 0$, but are
neutral, $Q=0$.

\medskip

The quantized charge-spin separated quantum numbers of the soliton excitations of the Peierls model \cite{SSH, ssh2, braz1} 
are recovered only
in the limit $m_{\up,\down} \to 0$. In this limit, 
the spin-dependent Dirac masses serve as regulators, and the result 
($Q=\pm 1, S^z=0$ or $Q=0, S^z = \pm 1/2$)
depends of
the relative sign of the vanishing amplitudes $m_{\up}$ and $m_{\rm \down}$.

\section{Antiadiabatic limit, quantum phonons. Bozonized Hamiltonian}\label{anti-adiabat}

We now turn to the field-theoretical description of
the Peierls model in the the anti-adiabatic limit, in which
the 
phonons are characterized by high frequency, 
$\omega_0 \gg |\Delta|, |m_{\s}|$.
This regime was first considered by Fradkin and Hirsch \cite{fh1}.
By integrating the phonons out, they derived a purely fermionic 
effective low-energy action which
has the form of a nonchiral N=2 Gross-Neveu (GN) 
model \cite{GN,witten1,shankarwitten,DHN,zamo2} 
with a local four-fermion interaction mediated by the phonons.
The staggered potentials appearing in (\ref{ion_pier_spinless}) supply the fermions with spin dependent
$\gamma^5$-masses. 
The resulting  fermionic model is described by the Lagrangian ${\cal L} = {\cal L}_{\rm GN} + 
{\cal L}_{\rm ext} $, where
\bea
{\cal L}_{\rm GN} &=& 
\ri \bar{\psi}_{\alpha} \gamma^{\mu}\p_{\mu}\psi_{\alpha}
+  \frac{1}{2} g^2 _0 \left(  \bar{\psi}_{\alpha} \psi_{\alpha} \right)^2 
\label{GN-L}\\
{\cal L}_{\rm ext} &=& - \ri \bar{\psi}_{\alpha}\gamma^5
\left( m_+ \delta_{\alpha\beta} + m_- \s^3 _{\alpha\beta} \right)
\psi_{\beta}
\label{ext}
\eea
Here 
$\alpha, \beta$ are the spin indices, $\bar{\psi} = \psi^{\dagger} \gamma^0$,
and for the Dirac $\gamma$-matrices the following representation is chosen:
$\gamma^0 = \htau_2$, $\gamma^1 = \ri \htau_1$, $\gamma^5 = \gamma^0 \gamma^1 = \htau_3$.

\subsection{Bosonization of O(4) Gross-Neveu model}\label{gn-bos}

For future purposes, using the bosonization method (see e.g. \cite{gnt}), 
we first recapitulate the 
main findings of Ref.\cite{fh1} that follow from 
the well-known properties of the massless
GN model (\ref{GN-L}). 
By
decoupling each complex fermion 
into a pair
of real  fields, 
one recasts the model (\ref{GN-L}) as an O(4)-invariant theory of
four interacting Majorana fermions 
\bea
{\cal H}_{\rm GN} 
= - \frac{\ri v_F}{2}~\xi^a \htau_3 \p_x \xi^a + \frac{1}{2} g^2 _0
\left( \xi^a _R \cdot \xi^a _L   \right)^2
\label{o4-maj}
\eea
where $\xi^a = (\xi^a)^{\dagger}~ (a=1,2,3,4)$.
Using the bosonization approach, one transforms
${\cal H}_{\rm GN}$ to a direct sum of two weakly perturbed
SU(2)$_1$ Wess-Zumino-Novikov-Witten (WZNW) models 
(see Appendix \ref{bosrules}) 
\bea
{\cal H}_{\rm GN} = {\cal H}_c + {\cal H}_s, ~~~ 
{\cal H}_{c,s} = {\cal H}^{(c,s)}_{\rm WZNW} + g^2 _0 {\bf J}^{c,s}_R \cdot {\bf J}^{c,s}_L,
~~~\label{pert-wzw}
\eea
where ${\bf J}^{c,s}_{R,L}$ are chiral vector currents of the corresponding critical
WZNW model (\ref{crit-wzw}). 
The charge-spin separated structure  of the Hamiltonian 
(\ref{pert-wzw}) reflects the symmetry group
equivalence  O(4) $\approx$ SU(2) $\otimes$ SU(2).
The theory (\ref{pert-wzw}) can be reformulated as
a sum of two
identical quantum sine-Gordon models 
for bosonic fields $\Phi_c$ and $\Phi_s$ (see Eq.(\ref{co2-expan})): 
\bea
{\cal H} (x) &=& \sum_{a=c,s}{\cal H}_{a}  (x), \label{2SG}\\
{\cal H}_{a} (x) &=& \frac{v}{2}\left\{ \left(  1 - \frac{g^2 _0}{2\pi v}\right)
\Pi^2 _{a} (x) + \left(  1 + \frac{g^2 _0}{2\pi v}\right)
\left( \p_x \Phi_{a} (x)  \right)^2
\right\} \nn\\
&& ~~-  \frac{g^2 _0}{2(\pi\alpha)^2} \cos \sqrt{8\pi} \Phi_{a} (x).
\label{SG}~~~~
\eea
All irrelevant corrections to Eq.(\ref{SG}) with the Gaussian scaling dimension
4 and higher are neglected (see however sec.\ref{1-comp}).
In terms of the original fermionic fields $\psi_{\alpha}$,
the cosine terms in (\ref{SG}) represent marginally relevant backscattering (spin sector)
and 
Umklapp processes (charge sector). 
On the Kosterlitz-Thouless phase diagram \cite{k},
each of the sine-Gordon Hamiltonians flows (in the RG sense) 
along the SU(2)-symmetric separatrix
towards the strong-coupling
infrared stable fixed point.

The main feature of the ground state of the GN model (\ref{o4-maj}) is a spontaneous breakdown
of the discrete $\gamma^5$-symmetry 
(with the continuous SO(4)-symmetry kept unbroken) and double degeneracy of the ground state.
The sum of the Majorana mass bilinears, $ \ri \sum_a \xi^a _R \xi^a _L  \sim  \sum_{\alpha}\bar{\psi}_{\alpha}\psi_{\alpha}$, acquires a nonzero vacuum expectation value \cite{witten1, shankarwitten}. 
The dynamically generated mass 
\be
m_c = m_s \equiv {\frm}
\simeq \Lambda \exp\left(  - \pi v_ /g^2 _0 \right) , ~~~ \Lambda \sim v_F / \alpha
\label{masses}
\ee
coincides with the single-soliton mass of the $\beta^2 = 8\pi$
SG model.
In terms of the original complex fermions the symmetry breaking order parameter is just the electron dimerization operator
(\ref{D}) which in the continuum limit has the following bosonic representation
(see Appendix \ref{bosrules})
\bea
D 
~\sim  \la {\rm Tr}~ \bar{\psi} \psi \ra ~\sim ~\cos \sqrt{2\pi} \Phi_c  \cos \sqrt{2\pi} \Phi_s 
\label{D-bosonized}
\eea
In the ground state, the fields $\Phi_c$ and $\Phi_s$ are locked at 
one of the degenerate minima of the cosine potential in (\ref{SG})
\bea
&&\left(  \Phi_c \right)_{j_c} = \sqrt{\frac{\pi}{2}} ~j_c, ~~~~
\left(  \Phi_s \right)_{j_s} = \sqrt{\frac{\pi}{2}} ~j_s, \label{cs-vac}\nn\\
&&(j_c , ~j_s= 0, \pm 1, \pm 2, \ldots)
\eea
The lattice of the PI vacua (\ref{cs-vac}) is shown in Fig. \ref{VACUA}(a).
Accordingly
\be
\la  D \ra_{j_c, j_s} = D_0 (-1)^{j_c + j_s}, \label{D-ave}
\ee
where (since the Gaussian scaling dimension of the dimerization operator is $d = 1$)
\bea
D_0 = |\la \cos \sqrt{2\pi} \Phi_c \ra  \la \cos \sqrt{2\pi} \Phi_s \ra |
= C^2  {\frm} / \Lambda \label{D0}
\eea
where $C$ is a numerical constant calculated in Ref.\cite{LZ1}.
Eq.(\ref{D-ave}) demonstrates double degeneracy of the SD 
ground state:
depending on the parity of the sum $j_c + j_s$, $\la D \ra$ takes values $\pm D_0$.

The spontaneously dimerized phase of the GN model is a typical example of
a strongly correlated state.
Contrary to the adiabatic model, a single-fermionic branch of the spectrum
of the model (\ref{2SG}), (\ref{SG}) is absent.
The only elementary excitations 
are topological quantum solitons of the 
$\beta^2 = 8\pi$ sine-Gordon model
(\ref{SG}), either in the charge or spin sector. They correspond to vacuum-vacuum transitions
with $\Delta \Phi_{c,s} = \pm \sqrt{\pi/2}$. 
These are the transitions between nearest-neighbor sites of the PI vacuum lattice, 
Fig. \ref{VACUA}(a).
According to the 
definitions (\ref{charge-spin}) given in Appendix \ref{bosrules},
the charge and spin carried by the solitons 
are defined as 
\bea
Q &=& \sqrt{\frac{2}{\pi}}\int_{-\infty}^{\infty} \rd x~\p_x \Phi_c (x)
= \sqrt{\frac{2}{\pi}} 
\Delta \Phi_{c}, \nn\\
S^z &=& \frac{1}{\sqrt{2\pi}}\int_{-\infty}^{\infty} \rd x~\p_x \Phi_{s}
= \frac{1}{\sqrt{2\pi}} \Delta \Phi_{s}
\label{charge-spin-bos}
\eea
and hence are equal to $Q = \pm 1$ and $~S^z = \pm 1/2$, respectively. Thus, 
the physical picture of a charge-spin separated solitonic spectrum of a 
commensurate Peierls insulator
emerging in the anti-adiabatic limit is qualitatively the same as
in the adiabatic model \cite{fh1}. In both limits spontaneous dimerization
of the system emerges at arbitrarily small electron-phonon coupling.
The SU(2) $\times$ SU(2) representation (\ref{2SG}), (\ref{SG}) of the O(4) GN model
proves efficient: the characterization of the spectrum in terms of the charge-spin separated
quantum solitons is in full agreement with the exact result \cite{zamo2, witten1} that the spectrum
of the O(4) GN model only consists of kinks.

\subsection{Bosonized form of staggered potentials and Band Insulator vacua}\label{bi-bos}

\begin{figure}[hbbp]
\centering
\includegraphics[width=3.8in]
{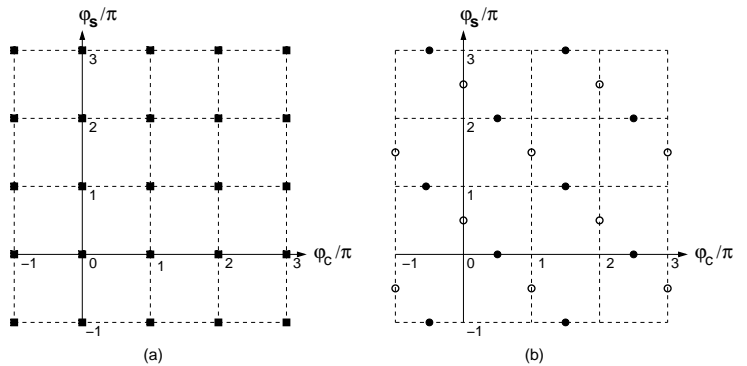}
\caption{\footnotesize Sets of vacua in PI and BI phases. Notations:
$\varphi_{c,s} = \sqrt{2\pi}\Phi_{c,s}$. (a) Lattice of PI vacua.
(b) BI vacua, black and open circles correspond to the cases $m_{+} > m_-$
and $m_+ < m_-$, respectively.}
\label{VACUA}
\end{figure}

Under the action of staggered one-particle potentials and in
the absence of electron-phonon coupling   ($g_0 = 0$)
the system represents a band insulator
with gapped single-particle excitations carrying both the charge and
spin quantum numbers. On the other hand, in a 
PI state 
single-particle excitations are absent
and the whole spectrum is exhausted by {collective} charge and spin modes described in terms
of quantum solitons. To understand the interplay between the phonon induced
electron-electron correlations 
that tend to dimerize the system and sign-alternating potentials that support a BI regime,
it is instrumental
to have a representation of the
staggered potentials in terms of scalar fields $\Phi_{c,s} = (\Phi_{\up} \pm \Phi_{\down})/\sqrt{2}$ describing collective charge and spin modes.
Using formulas (\ref{cdw-bos}) and (\ref{sdw-bos}) of Appendix \ref{bosrules},
we have
\bea
 \Delta H_{\rm stag} &=& \sum_{n\s} m_{\s} (-1)^n  c^{\dagger}_{n\s} c_{n\s}
\to \int \rd x ~{\cal H}_{\rm stag} (x); 
\nn\\
{\cal H}_{\rm stag}  = 
&-& \frac{1}{\pi\alpha} \left( m_{\up} \sin \sqrt{4\pi} \Phi_{\up} 
+ m_{\down} \sin \sqrt{4\pi} \Phi_{\down}  \right)\nn\\
 = &-& \left(\frac{2 m_+}{\pi \alpha}\right) \sin \sqrt{2\pi} \Phi_c  \cos \sqrt{2\pi} \Phi_s 
 \nn\\
&-& \left(\frac{2 m_-}{\pi \alpha}\right)  \cos \sqrt{2\pi} \Phi_c  \sin \sqrt{2\pi} \Phi_s .
\label{stag-bos}
\eea
(we remind that $m_{\pm} = (m_{\up} \pm m_{\down})/2$).
We see that, on one hand,
the BI Hamiltonian 
splits into 
spin-$\up$ and spin-$\down$ independent components, each of them representing the quantum sine-Gordon
model for the field $\Phi_{\s}$ with the coupling constant $\beta^2= 4\pi$ (free fermion point). 
Not surprisingly, no such decomposition is possible in terms of the charge and spin fields $\Phi_c$ and $\Phi_s$. The vacua of the BI phase are shown in Fig.\ref{VACUA}(b).

Without loss of generality everywhere below we will assume
that $m_{\pm} > 0$. Then from (\ref{stag-bos})
one determines the vacuum values of the fields $\Phi_c$ and $\Phi_s$:
\bea
\left(  \Phi \right)^{(c)}_{\rm vac} = \sqrt{\frac{\pi}{8}} + \sqrt{\frac{\pi}{2}} n_c, ~~
\left(  \Phi \right)^{(s)}_{\rm vac} = \sqrt{\frac{\pi}{2}} n_s
\label{vac-a}
\eea
at $m_+ > m_-$, and 
\bea
\left(  \Phi \right)^{(c)}_{\rm vac} = \sqrt{\frac{\pi}{2}} n_c,
~~\left(  \Phi \right)^{(s)}_{\rm vac} = \sqrt{\frac{\pi}{8}} + \sqrt{\frac{\pi}{2}} n_s,
\label{vac-b}
\eea
at $m_+ < m_-$, with the restriction that
the integers $n_c$ and $n_s$ have the same parity: $n_c + n_s
= 0, \pm 2, \pm 4, \ldots$. In such description, 
fundamental (single-particle) excitations in the noninteracting
BI are interpreted as topological excitations of the model (\ref{stag-bos})
associated with the transitions between nearest-neighbor degenerate vacua (\ref{vac-a}) or (\ref{vac-a}).
These are the transitions with
$|\Delta n_c| = |\Delta n_s| = 1$ which yield the charge $Q = \pm 1$ and
spin $S^z = \pm 1/2$.

\section{Anti-adiabatic Peierls model with staggered, spin-dependent potentials}

Having overviewed
the structure of the vacuum and elementary excitations 
for the "pure" PI ($m_{\pm} = 0$)
and BI ($g_0 = 0$) phases, here we consider
the general case with all perturbations to the Gaussian
part of the Hamiltonian  present.  
The total bosonized Hamiltonian reads:
\bea
{\cal H}(x) &=& {\cal H}^{(0)}_c + {\cal H}^{(0)}_s + U(\Phi_c, ~\Phi_s); \label{tot-ham}\\
U(\Phi_c, ~\Phi_s) =
&-& \frac{g^2 _0}{2(\pi\alpha)^2} \left( \cos \sqrt{8\pi} \Phi_c + \cos \sqrt{8\pi} \Phi_s   \right)\nn\\
&-&  \left(\frac{2m_+}{\pi \alpha}\right) \sin \sqrt{2\pi} \Phi_c \cos \sqrt{2\pi} \Phi_s 
\nn\\
&-& \left(\frac{2 m_-}{\pi \alpha}\right)  \cos \sqrt{2\pi} \Phi_c  \sin \sqrt{2\pi} \Phi_s, 
\label{ham:pi+stag}
\eea
where ${\cal H}^{(0)}_{c,s}$ are the Hamiltonians of free massless bosons in the charge and spin
sectors, given by the first line in Eq.(\ref{SG}). In this section we will study in detail
the interplay between the competing marginally relevant GN two-cosine perturbation
and strongly relevant staggered potentials and describe the outcome of this competition.

\subsection{Classical vacua} 

Stable vacua of the potential
$U(\Phi_c, \Phi_s)$ in Eq.(\ref{ham:pi+stag}) identify massive ground-state phases of the model.
A straightforward analysis done in Appendix \ref{minima} reveals the following properties of the potential.
In the regions $|m^2 _+  - m^2 _-| > \left( g^2 _0/\pi \alpha \right)^2$
the degenerate vacua of 
$U(\Phi_c, ~\Phi_s)$ 
form a square lattice with periods $\sqrt{\pi/2}$, Fig.\ref{VACUA}(b). These vacua are not affected by the electron-phonon
interaction and coincide with those of the
"pure" band insulator, Eqs. (\ref{vac-a}),(\ref{vac-b}).
This is the non-dimerized
BI phase of the model.

\begin{figure}[hbbp]
\centering
\includegraphics[width=2.4in]
{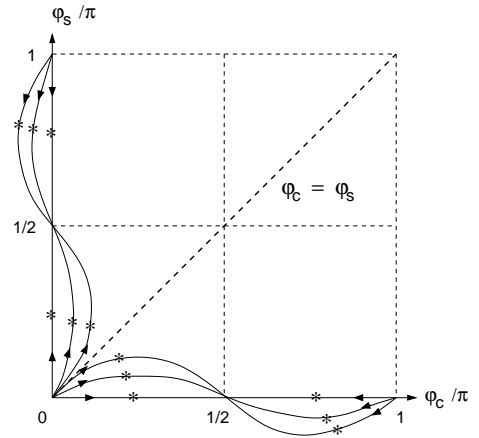}
\caption{\footnotesize 
Evolution of the minima  of the potential $U(\varphi_c, \varphi_s)$
(shown by the asterisks)
in the spontaneously dimerized phase $|m^2 _+ - m^2 _-| < (g^2 _0/\pi v_F)$
upon increasing $m_{\pm}$. Notations:
$\varphi_{c,s} = \sqrt{2\pi}\Phi_{c,s}$. }
\label{motion}
\end{figure}

At  $|m^2 _+  - m^2 _-| < \left( g^2 _0/\pi \alpha \right)^2$ the profile of the potential
changes. 
At $m_{\pm} = 0$
the SD massive phase
has a different set of vacua,
Eqs.(\ref{cs-vac}), Fig.\ref{VACUA}(a). 
As schematically shown in Fig.\ref{motion},
upon increasing the difference $|m^2 _+  - m^2 _-|$,
symmetrically located
pairs of new minima appear 
and start moving from their neighboring initial PI values towards the 
nearest BI value.
Depending 
on the ratio $m_+ / m_-$, the trajectories in the $(\Phi_c, \Phi_c)$ plane along which the minima of the potential 
$U$ 
move 
split in two groups.
The line $\Phi_c = \Phi_s$ is a separatrix which sets apart the CDW-dominated sector 
($m_+ > m_-$)
and the SDW-dominated sector ($m_+ < m_-$).
In the CDW sector 
the vacuum
values of the charge field
move 
from 
$\Phi_c = 0$ and $\sqrt{\pi/2}$ towards
the BI value $\Phi_c = \sqrt{\pi/8}$ where the two minima merge. The spin field 
$\Phi_s$ stays equal zero at both initial and final points of the trajectory
and satisfies the condition
$\Phi_s < \Phi_c$ in between (see Fig.\ref{motion}). 
The picture in the  SDW sector is similar and is obtained from the former case
by interchanging $m_+ \leftrightarrow m_-$,
$\Phi_c \leftrightarrow \Phi_s$.

\noindent

Thus, the equations
\be
m^2 _+  - m^2 _- = \pm \left( \frac{g^2 _0}{\pi \alpha}  \right)^2
\label{crit-lines}
\ee
determine classical critical lines separating phases with and without spontaneous dimerization. 
We will see below that quantum fluctuations significantly modify the equations for the critical curves.

Let us consider the region $|m^2 _+  - m^2 _- | < \left(  g^2 _0 /(\pi \alpha) \right)^2$
and inspect
the shape of  the potential in the vicinity of the critical lines (\ref{crit-lines}).
Below we will specialize to the case 
$m_+ > m_-$.
It is convenient to use the $(\Phi_{\up}, \Phi_{\down})$ representation of the potential
and pass
to dimensionless quantities. The details can be found in Appendix \ref{minima}.
The potential 
takes the form (see Eq.(\ref{POT})):
\bea
{\fU} (x,y) &\equiv& \left( \frac{\pi \alpha}{g_0}  \right)^2 U(\Phi_{\up}, \Phi_{\down})~~~\nn\\
&=& - \cos x \cos y 
 - a\sin x - b \sin y~~
\label{dimless-pot}
\eea
Let us expand ${\fU} (x,y)$ in the vicinity of the point $(x,y) = (\pi/2, \pi/2)$
which is a minimum of ${\fU}$ 
in the BI phase. 
For small deviations 
from this point, $q_1 = x - \pi/2$, $q_2 = y - \pi/2$, to the accuracy $O(q^6)$,
we have 
\bea
{\fU} (q_1, q_2) &=& {\fU} (0, 0) + \left[\frac{1}{2} \left( aq^2 _1 + b q^2 _2  \right) - q_1 q_2
\right]\nn\\
 &+&  \left[\frac{1}{6} (q^2 _1 + q^2 _2) q_1 q_2
- \frac{1}{24} \left( a q^4 _1 + b q^4 _2  \right) \right]
\label{U-expan-crit}
\eea
The quadratic part of ${\fU}$ is diagonalized by an appropriate
SO(2) rotation of the vector ${\bf q} = (q_1, q_2)$:
$\delta {\fU}^{(2)} = 
\frac{1}{2} \left(\lambda_1 \eta^2 _1 + \lambda_2 \eta^2 _2   \right)$,
where $\eta_1$ and $\eta_2$ are normal coordinates, and
\[
\lambda_{1,2} = f_+ \pm \sqrt{ f^2 _- + 1}, ~~~
f_{\pm} = (a \pm b)/2 = m_{\pm} \pi \alpha/g^2 _0.
\]
In the vicinity of the transition point we can use the parametrization
$f_+ = \left( 1 + \tau/2 \right) \cosh \gamma$, $f_- = 
\left( 1 + \tau/2 \right) \sinh \gamma$, so that $f^2 _+ - f^2 _- = 1 + \tau$,
where $\gamma > 0$ 
and $|\tau| \ll 1$ such that 
$\tau > 0$ ($\tau < 0$) for the BI (PI) phase.
So
\bea
\lambda_1 
= 2\cosh \gamma + O(\tau), ~~~~
\lambda_2 = \frac{\tau}{2\cosh \gamma} \label{l12}
\eea
From (\ref{l12}) it follows that the point $(x,y) = (\pi/2, \pi/2)$
is a minimum of the potential at $\tau > 0$ (BI phase) and a saddle point at $\tau <0$
(SD phase). The rigidities of the potential around this point in the $\eta_1$ and $\eta_2$
directions are drastically different: $\eta_2$ is a soft direction where the stiffness
$\lambda_2 \sim \tau$ is small,  whereas in the orthogonal direction $\eta_1$ the stiffness is much larger,
$\lambda_1 \geq 2$. So, in the vicinity of the point $(\pi/2, \pi/2)$, at small deviations from
the criticality ($|\tau| \ll 1$) the potential has the shape of a narrow channel 
along the $\eta_2$-direction.
For this reason, in the low-energy limit, it is sufficient to consider the effective potential
${\fU} (\eta_1, \eta_2)$ only along the "easy" direction $\eta_2$. 
The resulting expansion of the potential 
\bea
{\fU} (\eta_1, \eta_2) &\to& {\fU} (0, \eta_2)~~~~~\nn\\
&=& {\rm const}~+~ \frac{\tau}{4\cosh \alpha}~\eta^2 _2 ~+~ \frac{1}{8\cosh \alpha}~\eta^4 _2
~~~\label{doub-well-pot}
\eea
reveals its effectively one-dimensional double-well structure which unambiguously indicates
the existence of an Ising transition at $\tau = 0$.

\subsection{Quantum approach}

\subsubsection{Perturbative estimations}

Having identified stable vacua of the potential $U(\Phi_c, \Phi_s)$,
now we turn to a quantum description of the 
PI-to-BI crossover. 
According to (\ref{cs-vac}), in the unperturbed PI phase ($m_{\pm} = 0$)
the vacuum values of the fields $\Phi_c$ and $\Phi_s$ are 
multiples of $\sqrt{\pi/2}$,  so that the averages 
\be
|\la \cos \sqrt{2\pi} \Phi_{c,s} (0) \ra_0| = C_0 ({\frm}/\Lambda)^{1/2} \neq 0
\label{1p-ave1}
\ee
(the numerical value of the prefactor $C_0$ is calculated in Ref.\cite{LZ1}).
Consider the regime
$m_{\pm}\ll{\frm}$, 
in which the staggered fields can be treated as small perturbations.
Apparently, perturbative expansions of $\la \cos \sqrt{2\pi} \Phi_a (0) \ra ~(a=1,2)$ contain only even powers of the staggered amplitudes $m_{\pm}$:
\bea
&&\la \cos \sqrt{2\pi} \Phi_{c,s} (0) \ra \nn\\
 &&=  C_0 \left( {\frm}/{\Lambda}  \right)^{1/2}
\left[ 1 - C
\left( {m_{\pm}}/{\frm}  \right)^2 + \cdots
\right]
\label{cos-ave}
\eea
Here $C \sim 1$ is a positive numerical constant.
We observe
that perturbation theory breaks down at $m_{\pm} \sim {\frm}$,
which is roughly
the conditions for the Ising criticalities.

\subsubsection{Reduction to coupled double-frequency sine-Gordon models}

In the pure PI phase  
the operators $\cos \sqrt{2\pi} \Phi_{c,s}$ have  
nonzero expectation values, whereas
$\sin \sqrt{2\pi} \Phi_{c,s}$
are short-ranged fluctuating fields with zero averages.
Therefore,  in the leading order, the staggered potentials in (\ref{ham:pi+stag})
can be replaced
by
\bea
{\cal H}_{\rm stag} \to- \left(\frac{2\bar{m}_+}{\pi \alpha}\right) \sin \sqrt{2\pi} \Phi_c  
- \left(\frac{2 \bar{m}_-}{\pi \alpha}\right)   \sin \sqrt{2\pi} \Phi_s ~~~~~~
\label{eff-stag}
\eea
with
\be
\bar{m}_{+} = m_{+}  \la \cos \sqrt{2\pi} \Phi_{s} \ra, ~~~~
\bar{m}_{-} = m_{-}  \la \cos \sqrt{2\pi} \Phi_{c} \ra
\label{eff-ampl}
\ee
As a result, the effective Hamiltonian decouples into two quantum \emph{double-frequency sine-Gordon} (DSG) models
\bea
&& {\cal H}_{\rm eff} = \sum_{a=c,s}{\cal H}^{(a)}_{\rm DSG}, \label{additive}\\
&& {\cal H}^{(a)}_{\rm DSG} = {\cal H}^{(a)}_0 - \eta_a 
\cos \sqrt{8\pi} \Phi_{a} - \zeta_{a} \sin \sqrt{2\pi} \Phi_{a}
\label{dsg-model}\\
&& \eta_c =\eta_s = \frac{g^2 _0}{2(\pi\alpha)^2}, ~~\zeta_c = \frac{2\bar{m}_+}{\pi\alpha},
~~\zeta_s = \frac{2\bar{m}_-}{\pi \alpha}~~~~
\label{dsg-param}
\eea
The DSG model has been analyzed in detail in Refs. [\onlinecite{DM},\onlinecite{FGN2}].
It describes an interplay between two relevant perturbations to the Gaussian conformal field theory
${\cal H}^{(a)}_0$
with the ratio of their scaling dimensions equal to 4. Because the two perturbations have different
parity symmetries and, consequently, the field configurations which minimize one perturbation do not minimize the other, the competition between them produces an Ising \emph{quantum} critical point \cite{DM}.

The additive structure of the Hamiltonian (\ref{additive}) might lead to a wrong
conclusion that the effective model has a charge-spin separated form and hence
should display two independent quantum Ising transitions, one in each sector. 
In fact, in agreement with
the above classical analysis, at a given ratio
$m_+ / m_- \neq 1$,
there exists only one transition. At the quantum level, the key point is the self-consistency
conditions (\ref{eff-ampl}) that couple the charge and spin sectors. We will argue now that, in the
CDW-like case ($m_+ > m_-$) the Ising transition is described by ${\cal H}^{(c)}_{\rm DSG}$,
whereas the criticality in ${\cal H}^{(s)}_{\rm DSG}$ is avoided. In the SDW-like case 
($m_+ < m_-$)
the situation is just the opposite.

The critical point in the DSG model (\ref{dsg-model}) can be estimated by comparing the mass gaps that would open up if the two perturbations were acting alone. The $\eta_a$-term generates
the Peierls mass gap ${\frm}$, Eq.(\ref{masses}).  The mass gap generated by the
$\zeta_a$-term scales as
$
m_{\zeta_a}  \sim \Lambda \left(  \zeta_a \alpha /\Lambda \right)^{2/3}.
$
So the critical values of the effective staggered amplitudes,
$\bar{m}_{\pm}^*$, should coincide and be of the order of
\be
\bar{m}_{+}^* =  \bar{m}_{-}^* \sim \Lambda \left(  {\frm}/\Lambda \right)^{3/2} 
\label{crit-points}
\ee
Exponential smallness of the r.h.s. of (\ref{crit-points}) shows that
quantum fluctuations significantly reduce the critical value of the staggered amplitudes
as compared to the classical estimate (\ref{crit-lines}).

If for any bare values of $m_{\pm}$  the transitions in the charge and spin DSG models
(\ref{dsg-model}) had occurred independently, then
Eqs.(\ref{crit-points}) and (\ref{eff-ampl}) would imply that 
$
m_+ / m_- = {\la \cos \sqrt{2\pi} \Phi_c \ra}/{\la \cos \sqrt{2\pi} \Phi_s \ra}.
$
However, already the perturbative expansions (\ref{cos-ave})  show that the above relation cannot be valid except for the
special case $m_+ = m_-$ (see below).
Therefore, at a given ratio
$ m_+ / m_- \neq 1$, there can be only one Ising transition: either in the charge DSG model
if $m_+ > m_-$,  or in the spin DSG model if $m_+ < m_-$.

It has been shown in Ref.[\onlinecite{FGN2}] that mapping of the DSG model (\ref{dsg-model})
onto a generalized 1D quantum Ashkin-Teller model makes the Ising criticality accessible by
non-perturbative means. 
The lowest-energy sector of the theory is described by
a single critical quantum Ising model, i.e. a CFT with central charge $c=1/2$.
Using this correspondence, one can find the low-energy projections of
the physical fields near the transition. In particular, it has been shown \cite{FGN2} that
$
\cos\sqrt{2\pi} \Phi_a \sim\mu$, $\sin \sqrt{2\pi} \Phi_a \sim I + \vare$,
where $\mu$ and $\vare$ are the disorder field and energy density of the critical Ising 
model \cite{cft}, $I$ being the identity operator. Accordingly,
the average values of these operators are

\bea
&&\la \cos\sqrt{2\pi} \Phi_a \ra_{\zeta_a} \sim \left[\frac{(\zeta_a ^* - \zeta_a } 
{\zeta_a ^*}  \right]^{1/8}, ~~\zeta_a  < \zeta_a ^* \nn\\
&& ~~~~~~~~~~~~~~~~~~=~~~~0, ~~~~~~~~~~~ ~~~~~~\zeta_a  > \zeta_a ^*  \label{ave1}\\
&& \la \sin \sqrt{2\pi} \Phi_a \ra_{\zeta_a} -  \la   \sin \sqrt{2\pi} \Phi_c\ra_{\zeta_a ^*}
\nn\\
&&~~~~~~~~~~~~~~~~~~~~
\sim~ \frac{\zeta_a - \zeta_a ^*}{\zeta_a ^*}
\ln \frac{\zeta_a ^*}{|\zeta_a - \zeta_a ^*|}
\label{ave2}
\eea
So, as follows from (\ref{ave1}), at $m_+ >m_-$, in the vicinity of the Ising transition
the effective staggered magnetic field $\bar{m}_- $ vanishes 
as $m_+ \to m^* _+$,
so that the condition ${\bar{m}_-} \sim \Lambda ({\frm}/\Lambda)^{3/2}$ cannot be satisfied.
On approaching the critical point
the spin degrees of freedom remain gapped since the DSG Hamiltonian ${\cal H}^{(s)}_{\rm DSG}$
effectively transforms to a SG model. 
The Ising critical point in the charge DSG model ${\cal H}^{(c)}_{\rm DSG}$ separates the mixed
massive phase in with coexisting site-diagonal CDW and SD ($m_+ < m^* +$) from the pure CDW phase where
dimerization vanishes ($m_+ > m^* _+$). 

Close to the transition the dimerization order parameter
(\ref{D-bosonized}) is proportional to $\la \cos \sqrt{2\pi} \Phi_c\ra$
and, according to (\ref{ave1}), vanishes as 
\be
\la D\ra \sim \Theta (m^* _+ - m_+)\left(m^* _+ - m_+\right)^{1/8}
\label{D-tran}
\ee
as $m_+ \to m_+ ^*$.
The singular part of the staggered charge\\ density $\la  \rho_{\rm stag} \ra$ is 
proportional to the average $\la \sin \sqrt{2\pi} \Phi_c \ra$. As follows from (\ref{ave2}),
it remains continuous across the transition, but its derivative which determines the staggered
compressibility of the system, displays a logarithmic singularity:
\be
\kappa_{\rm stag} = \p \la  \rho_{\rm stag} \ra/ \p m_+ 
\sim - \ln |m_+ - m^* _+|
\label{kappa}
\ee

Since the staggered potentials
break the SU(2)$\times$ SU(2) symmetry of the GN model,
on lowering the energy scale  in the charge sector the parameters of the Hamiltonian
${\cal H}_{\rm DSG}^{(c)}$, $\zeta_c$
and  $\eta_c$, will undergo renormalization. Moreover,
the compactification
radius of the field $\Phi_c$ (i.e. the Luttinger liquid parameter $K_c$ whose unperturbed value
is $K^{(0)}_c = \beta^2/8\pi = 1$) will also acquire a (nonuniversal) dependence
on $m^2 _{\pm}$. However, up to $O(m^2 _-)$ corrections,
the condition $\bar{m}^* _+  \sim \Lambda ({\frm}/\Lambda)^{3/2}$ implies that
in the region $m_+ > m_-$ the Ising transition occurs at
$
m_+ = m^* _+ \sim {\frm}.
$
In  the SDW-like case, $m_+ < m_-$, the situation is identical to the above scenario
with the charge and spin DSG models interchanged. Here the renormalized scalar amplitude
$\bar{m}^* _+$ vanishes at the Ising criticality, and the critical point is determined by the condition
$
m_- = m^* _- \sim {\frm}.
$
Formula (\ref{D-tran}) still holds with the replacement of $m_+$ by $m_-$, and
the logarithmic singularity at the transition becomes the property of the staggered
spin susceptibility of the system:
\be
\chi_{\rm stag} = \p \la  {s}^z_{\rm stag} \ra/ \p m_-
\sim - \ln |m_- - m^* _-| \label{chi-stag}
\ee

A special situation arises when $m_+  = \pm m_-$. This is the case when the staggered potential
acts only on fermions with a certain spin projection. This case will be considered
in sec.\ref{1-comp}. 
Here we only mention that this is the situation when  the two DSG models 
${\cal H}^{(c)}_{\rm DSG}$
and ${\cal H}^{(s)}_{\rm DSG}$
become identical, implying that the Ising criticality will be reached simultaneously in both sectors.
However, the universality class of the critical point will change.
The effective low-energy theory will be described 
in terms of  two copies of identical
critical quantum Ising models, or equivalently two species of massless Majorana fermions.
Since two Majorana fermions can be combined into a single Dirac (i.e. complex) fermion, the 
original discrete $\mathbb{Z}_2 \times \mathbb{Z}_2$ symmetry gets enlarged to the continuous $U(1)$.

\subsection{Topological excitations with fractional quantum numbers}

For the SD 
(mixed) phase realized at 
$|m^2 _+ - m^2 _- |^2 < (g^2 _0 /\pi\alpha)^2$, the set of degenerate vacua of the potential $U(\Phi_c, \Phi_s)$
is discussed in detail in Appendix \ref{minima}. 
In accordance with the double degeneracy of the SD ground state,
there exist two sets of minima $\vvphi = (\varphi_c, \varphi_s)$ [see  Eqs. (\ref{roots})]:
\bea
&& \vvphi^{(1)}_{>} = [\varphi_+ + \pi n_c, ~\varphi_- + \pi n_s],\nn\\
&& \vvphi^{(2)}_{>} = [- \varphi_+ + \pi (n_c + 1), ~- \varphi_- + \pi n_s]
\label{vacua-1}
\eea
at $m_+ > m_-$, and 
\bea
&& \vvphi^{(1)}_{<} = [\varphi_- + \pi n_c, ~\varphi_+ + \pi n_s],\nn\\
&& \vvphi^{(2)}_{<} = [- \varphi_- + \pi n_c, ~- \varphi_+ + \pi (n_s + 1)]
\label{vacua-2}
\eea
at $m_+ < m_-$. In the above formulas 
$\varphi_{c,s} = \sqrt{2\pi} \Phi_{c,s}$,
$\varphi_{\pm} = (x_0 \pm y_0)/2$, where $x_0$ and $y_0$ are given by the expressions (\ref{roots}). The parameters $\varphi_{\pm}$
vary within the interval $(0,\pi/2)$,
and 
the sum of integers $n_c$ and $n_s$ is constrained to be an even number.
\begin{figure}[hbbp]
\centering
\includegraphics[width=2.5in]
{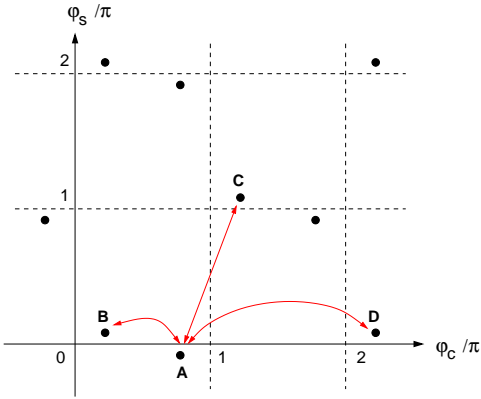}
\caption{\footnotesize The set of degenerate vacua of the potential
$U(\Phi_c, \Phi_s)$ in the mixed SD phase at $m_+ > m_-$.
The coordinates of the points A,B,C,D are
$A = (\pi - \varphi_+, - \varphi_-)$, $B = (\varphi_+, ~\varphi_-)$,
$ C = (\pi + \varphi_+, ~\pi - \varphi_-)$, $D^{(1)} = (2\pi + \varphi_+, ~\varphi_-)$.
}
\label{SD-vacua-phase}
\end{figure}

The location of the vacua in the CDW-like phase ($m_+ > m_-$) is shown
Fig.\ref{SD-vacua-phase}.
Stable topological excitations represent dimerization kinks which
correspond to 
$\vvphi^{(1)} \leftrightarrow \vvphi^{(2)}$ transitions between nearest vacua.
These are of the AB, AC and AD types. 
The charge $Q$ and spin $S^z$ 
of these excitations, 
determined by formulas (\ref{charge-spin-bos}), take fractional (non-quantized) values
because at any nonzero $m_+ \neq m_- $ the SU(2) $\times$ SU(2) symmetry of the pure
PI phase is explicitly broken. 
The double-well periodic structure of the potential,
typical for the DSG model \cite{DM},
implies the existence of
"short" (AB) and "long" (AD) kinks carrying the
quantum numbers 
\be
Q_{\mp} = 1 \mp \frac{2\varphi_+}{\pi}, ~~~
S^z = \mp \frac{\varphi_-}{\pi} \label{short-long}
\ee
In the limit $\varphi_{\pm} \to 0$ 
both kinks 
convert to a spinless soliton
of the pure PI model: $Q_{\pm} \to 1$, $S^z \to 0$. On approaching the SD-BI transition point
($\varphi_+ \to \pi/2$, $\varphi_- \to 0$) the AD and AB kinks transform to
spin-singlet states ($S^z _{\pm} = 0$) either of
two particles (holes),  $Q_+ = 2$, or a particle-hole state, $Q_- = 0$.
The AD transition describes an excitation with the spin $S^z = 1/2$ and the charge
$Q = {2\varphi_+}/{\pi}$, interpolating
between a neutral spin-1/2 soliton 
of the PI and
a quasiparticle of the BI. 
The AD and AC excitations stay massive
across the Ising transition. However, at criticality, the short kink AB loses its charge
and mass and transforms to a collective gapless excitonic mode (c.f. Ref.\cite{FGN1}).

Comparing (\ref{vacua-1}) and (\ref{vacua-2}) one finds out that the quantum numbers of topological excitations in the SDW-like phase ($m_+ < m_-$)
can be deduced from the previous case
by using the interchange symmetry $\varphi_{+} \leftrightarrow \varphi_-$,
$Q \leftrightarrow 2S^z$. The analog of the short and long kinks (\ref{short-long})
is
\bea
Q = \pm \frac{2\varphi_+}{\pi}, ~~~S^z = \frac{1}{2} \mp \frac{\varphi_-}{\pi}
\label{short-long1}
\eea
These kinks interpolate between a neutral spin-1/2 soliton of the pure PI model at
$\varphi_{\pm} \to 0$ and particle-hole states with the total spin $S=0$ or $1$ at
the transition point ($\varphi_+ \to 0$, $\varphi_- \to \pi/2$). There is also an excitation
with the charge $Q=1$ and fractional spin $S^z = {\varphi_-}/{\pi}$ which interpolates between
a spinless soliton carrying a unit of charge in the pure PI phase and a quasiparticle
of the BI insulator. As in the CDW-like phase, here too we find that at the criticality
there exists a neutral collective mode that loses its mass and has zero spin.

\begin{figure}[hbbp]
\centering
\includegraphics[width=2.6in]
{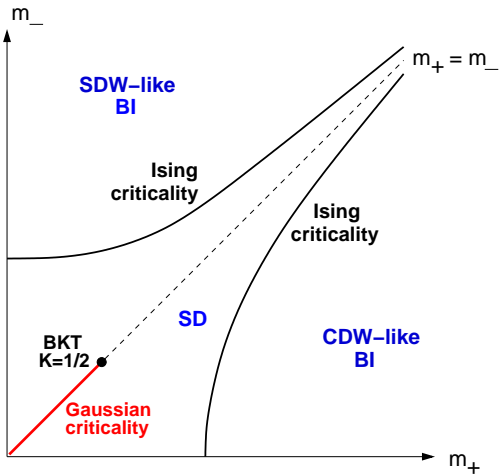}
\caption{\footnotesize Phase diagram of the model in the anti-adiabatic limit..}
\label{antiadiabatic-phase-dia}
\end{figure}

\section
{Peierls model in the vicinity of the line ${\bf m_+ = m_-}$}
\label{1-comp}

In this section we consider  the situation when
the amplitudes of the staggered potentials $m_{\up,\down}$
are strongly different, e.g.
$
m_{\up} \gg m_{\down}.
$ 
As before, it will be still assumed that $0<m_{\up,\down} \ll \Lambda$.
The above condition 
includes the case when the sign-alternating potential
is applied to one spin component only, e.g. 
\be
m_{\down} = 0, ~~~m_+ = m_-
\label{down-0}
\ee

Let us first consider the  case (\ref{down-0}) in the adiabatic limit.
A nontrivial solution for the order parameter,
\be
\Delta^2 = - \frac{m_{\up}}{2} + \sqrt{\frac{m^2 _{\up}}{4} + \Delta^4 _0},
\label{Delta-asym}
\ee
indicates the onset of spontaneous
dimerization
at {arbitrarily small} electron-phonon coupling $\lambda_0$
for any nonzero $m_{\up}$. This fact is entirely due
to the existence of a gapless fermionic component (with $\s = \up$)
and the adiabatic approximation we adopted here. As follows from (\ref{Delta-asym}),
when the electron-phonon interaction is strong enough, 
$ m_{\up} \ll  \Delta_0$, the massiveness of the
spin-$\up$ fermions is unimportant, and, in the leading order, $\Delta$ is given by
the standard expression for a canonical (massless) Peierls model,
$\Delta \simeq \Delta_0$. For a weak 
electron-phonon coupling, $m_{\up} \gg \Delta_0$, $\Delta$ vanishes at $\Delta_0 \to 0$ according
to the quadratic law $\Delta \simeq \Delta^2 _0 / m_{\up}$.

Now we turn to the anti-adiabatic limit of quantum phonons and consider the case
$m_{\up} \gg m_{\down}$.
Starting from the massive GN model
with spin-dependent Dirac masses 
\bea
{\cal H} &=& \sum_{\s} 
\psi^{\dagger}_{\s} \left(- \ri v_F \htau_3 \p_x + m_{\s} \htau_1   \right)\psi_{\s}
\nn\\
&&~~-~   \frac{1}{2} g^2 _0\left( \sum_{\s}\psi^{\dagger}_{\s} \htau_2 \psi_{\s} \right)^2
\label{GN-new}
\eea
we will integrate the heavy spin-$\up$ electrons
out and derive an effective Hamiltonian describing the low-energy degrees of freedom for
the light $\s=\down$ electrons.

Apart from the usual Umklapp processes 
$
R^{\dagger}_{\up} R^{\dagger}_{\down} L_{\down} L_{\up} + h.c.
 \sim \cos\sqrt{8\pi} \Phi_c
$
involving electrons with opposite spin projections, the interaction in (\ref{GN-new})
also includes Umklapp scattering of the electrons with the same spin projection:
\bea
&& \sum_{\s} \left [R^{\dagger}_{\s} (x) R^{\dagger}_{\s} (x + \alpha)L_{\s} (x + \alpha)
L_{\s} (x)  + h.c. \right]\nn\\ 
&&~~~~~~~~~~~~~~~~~~~~~~~~ \sim ~ \sum_{\s} \cos \sqrt{16 \pi} \Phi_{\s} (x)
\label{new-umkl}
\eea
At the Gaussian ultraviolet fixed point such perturbation has scaling dimension 4
and hence  is strongly irrelevant.
However, the staggered potentials break the SU(2)$\times$SU(2) symmetry of the GN model
and cause renormalization of the compactification radii of the scalar fields. 
If this renormalization is strong enough,
the Umklapp processes (\ref{new-umkl}) may become relevant and thus affect the phase diagram of the system. Such situation is known to exist in 
a 1D half-filled tight-binding model of spinless fermions with a strong enough 
nearest-neighbor repulsion (by the Jordan-Wigner
correspondence, this is equivalent
to a XXZ spin-1/2 chain with exchange anisotropy parameter $J_z/J_x > 1$) \cite{giam}.

We will assume that $m_{\up}$ is large enough to suppress
the Umklapp and forward scattering processes 
in the spin-$\up$ component.
So the "heavy" subsystem can be treated in terms of free massive fermions with
the mass $m_{\up}$.
The unperturbed Euclidian action for the spin-$\down$ electrons 
has the form of a DSG model
\bea
&&S^{(0)}_{\down} = \int \rd^2 {\bf r}~
\Big\{
\frac{1}{2} \left(1 + \frac{g^2 _0}{4\pi v_F}\right)
\left( \vnab \Phi_{\down}({\bf r}) \right)^2
\nn\\ 
&&- \frac{m_{\down}}{\pi\alpha v_F} \sin \sqrt{4\pi} \Phi_{\down}({\bf r})- 
\frac{g^2 _0}
{(2\pi\alpha)^2 v_F} \cos \sqrt{16 \pi} \Phi_{\down}({\bf r})\Big\}~~~
\label{S0-down}
\eea
where ${\bf r} = (v_F \tau, x)$ is a 2D Euclidian coordinate.
For a weak electron-phonon coupling, the last term in (\ref{S0-down}) remains strongly
irrelevant, with the scaling dimension
$
d = 4 \left( 1 - g^2 _0 /4\pi v_F + \cdots\right),
$
still close to 4.
In such case (\ref{S0-down}) reduces to the $\beta^2 = 4\pi K$ SG model describing massive spinless fermions
with a weak short-distance repulsion (massive Thirring model \cite{gnt}).
As shown in Appendix \ref{inter},
interaction between the "light" (spin-$\down$) and "heavy" (spin-$\up$)
fermions leads to a significant renormalization of the parameters of the effective action
$S_{\down}$. The second-order correction $\Delta S_{\down}$ to the "bare"
action $S^{(0)}_{\down}$ is given by formula (\ref{correction1}).
Adding it to (\ref{S0-down})
and rescaling the field,
$
\Phi_{\down} \to \sqrt{K} \Phi_{\down}
$
we arrive at
the effective 
action 
\bea
&&S^{({\rm eff})}_{\down} = S^{(0)}_{\down} + \Delta S^{({\rm eff})}_{\down}
= \int \rd^2 {\bf x}~
\Big\{
\frac{1}{2} (\vnab \Phi_{\down})^2 \nn\\
&& - \frac{\eta}{\alpha^2} \sin \sqrt{4\pi K} \Phi_{\down}
- \frac{\zeta}{(2\pi \alpha)^2} \cos \sqrt{16 \pi K} \Phi_{\down}
\Big\}
\label{eff-action-down}
\eea
where
\be
\eta = \frac{m_{\down} \alpha}{\pi v_F}, ~~~~
\zeta = \frac{g^2 _0}{v_F} + C_1 \left(   \frac{g^2 _0}{v_F} \right)^2 
\left( \frac{\Lambda}{m_{\up}}  \right)^2
\ee
are dimensionless amplitudes and
\be
\frac{1}{K} = 1 + \frac{g^2 _0}{4\pi v_F} + 
\frac{1}{2}\left( \frac{g^2 _0}{\pi v_F}  \right)^2 \ln \frac{\Lambda}{m_{\up}}
\label{K}
\ee
We see that renormalization effects arising in the second order in $(g_0 ^2)$ due to 
the coupling between the electrons with opposite spins enhance the values of
the parameters $K^{-1}$ and $\zeta$.

Let us first consider the case $m_{\down} = 0 ~(\eta=0)$.
The action (\ref{eff-action-down})
is a SG model at $\beta^2 = 16 \pi K$
and represents a bosonized version of the massless N=1 GN model \cite{fh1}.
Even though the dependence
of the parameters $K$ and $\zeta$ 
on the bare coupling constants and
the mass $m_{\up}$ are nonuniversal, Eq.(\ref{K}) suggests 
that in the region $(g^2 _0 /v_F)^2 \ln (\Lambda/m_{\up})\sim 1$
a critical value of $K=K_c = 1/2$ should be reached, below which the Umklapp term in 
(\ref{eff-action-down}) is
relevant and supports spontaneous dimerization of the spin-$\down$ electron subsystem.
Thus, at $K > 1/2$ the cosine perturbation is strongly irrelevant, and the spin-$\down$ fermionic subsystem is in a Luttinger-liquid gapless phase \cite{gnt,giam}
with critical exponents continuously varying with $K$. The line of Gaussian fixed points terminates at $K_c = 1/2$ (see Fig.\ref{antiadiabatic-phase-dia}). At this point the 
system undergoes a BKT 
transition to a spinless massive Peierls phase. 
The critical value of the "heavy" component mass $m_{\up}$ when the Peierls transition occurs is
extremely small
\be
m_{\up} \simeq \Lambda \exp \left[ - \frac{1}{2} \left( \frac{2\pi v_F}{g^2 _0}  \right)^2  \right]
\label{crit-m-up}
\ee

We see that, in the case when a staggered field is applied only
to one spin component, the picture emerging in the adiabatic approach differs
from that emerging in the anti-adiabatic limit: in the former case the onset of SD occurs
at arbitrarily small electron-phonon coupling, whereas in the latter case there exists a
BKT transition from the metallic phase to the PI phase of spin-$\down$ fermions.

When the amplitude $m_{\down}$ is finite but still much less than $m_{\up}$,
the effective action (\ref{eff-action-down}) contains the
$\eta$-term which supports a BI phase. The negative sign of the Umklapp amplitude ensures that at any nonzero $m_{\down}$
there should exist an Ising criticality at some critical value
of $m_{\down} = m^{(\rm crit)}_{\down} (g_0 ^2, m_{\up})$ separating the SD and BI massive phases.

Fig.\ref{antiadiabatic-phase-dia} summarizes the main qualitative features 
of the anti-adiabatic model.

\section{Conclusions}

In this paper, we have studied the ground-state 
properties of the one-dimensional
Peierls model  at 1/2-filling \cite{SSH, ssh2} subjected to a simultaneous action of
staggered scalar potential ($m_+$) and a sign-alternating magnetic field ($m_-$).
Our goal was to describe the interplay between the electron-phonon interaction which, when
acting alone, would
lead to the onset of a spontaneously dimerized Peierls 
phase with a charge-spin separated low-energy 
spectrum, 
and external, spin-dependent,
staggered single-particle potentials that transform a one-dimensional metal to a 
standard band insulator with CDW and SDW superstructures and the usual
quasiparticle excitations, carrying both the charge and spin.
The electron-phonon interaction has been treated using both the semiclassical, adiabatic approximation and in the
anti-adiabatic limit in which the high-frequency quantum phonons generate an effective
attraction between the electrons. We specialized to the weak coupling regime in which the
low-energy degrees of freedom are adequately described by the continuum electron-phonon model
\cite{tll} with spin-dependent Dirac masses $m_{\up,\down} = (m_+ \pm m_-)/2$.

The 
phase diagram of the model 
is depicted in Fig.\ref{antiadiabatic-phase-dia}. 
Depending on the sign of the parameter $F = |m^2 _+ - m^2 _-|> (g^2 _0 / \pi \alpha)^2$,
where $g_0$ and $\alpha$ are the electron-phonon coupling constant and short-distance
cutoff parameter, respectively,
the system occurs in one of three gapped phases: a CDW dominated band insulator phase
realized 
at $F > 0$,  $m_+ > m_-$, a SDW dominated band insulator
at $F > 0$, $m_+ < m_-$, and a mixed Peierls phase at $F<0$, in which the CDW and SDW superstructures
coexist with a nonzero spontaneous dimerization.
The massive phases are separated by critical lines ($F=0$) belonging to the universality class of 
the quantum Ising model. 
Except for the symmetry line $m_+ = m_-$, quantum effects do not
change the phase diagram qualitatively, i.e. only affect the location of the critical lines.

The semiclassical treatment of the phonon degrees of freedom
shows that at 
$m_{\pm} \neq 0$ spontaneous dimerization of the system occurs 
at a finite critical value of the dimensionless electron-phonon coupling constant,
$\lambda_c \equiv \left( 2 g^2 _0 /\pi v_F \right)_c = \ln (2W/\sqrt{|m_{\up} m_{\down}|})$, 
$W$ being of the order of the electron bandwidth.
The onset of the Ising criticality
is accompanied by the Kohn anomaly in the renormalized phonon spectrum.
We have demonstrated that the adiabatic approximation cannot provide a
satisfactory description of the model in the vicinity of the Ising criticality.
We have derived a Ginzburg criterion which determines a narrow region
around the critical point, where quantum fluctuations play a dominant role and 
the adiabatic (or mean-field) approximation is no longer applicable.

The  line 
$m_+ = m_-$ is a special case of an external staggered potential applied
to one fermionic spin component only. In the adiabatic approximation, along this line
the system is unstable against a spontaneous dimerization at arbitrarily
small electron-phonon coupling. We show that, contrary to such conclusion, the
analysis carried out in the anti-adiabatic limit, indicates the existence of
a Berezinskii-Kosterlitz-Thouless critical point  which separates a Luttinger-liquid gapless phase from the spontaneously dimerized one (see Fig.\ref{antiadiabatic-phase-dia}).

In the anti-adiabatic limit, where the quantum $\pi$-phonons are characterized by
a high frequency $\omega_0 \gg \Delta_0 / \sqrt{\lambda_0}$ ($\Delta_0$ being the
electron  spectral gap of the Peierls insulator),
the effective low-energy model represents a \emph{massive} version of the non-chiral N=2 Gross-Neveu model which, apart from the conventional four-fermion interaction induced by
the phonons \cite{fh1}, also includes spin-dependent fermionic $\gamma^5$-mass terms. 
We bosonized this model 
in terms of two 
scalar fields,
$\Phi_c$ and $\Phi_s$, describing the collective charge and spin degrees of freedom.
The resulting continuous model is given by Eq.(\ref{tot-ham}) and represents a charge-spin separated sum of two $\beta^2 = 8\pi$
sine-Gordon models coupled by  strongly relevant $m_{\pm}$-perturbations.
For all $m_+ \neq m_-$ the phase diagram of 
Fig.\ref{antiadiabatic-phase-dia}
was derived from the analysis of stable vacua of the potential (\ref{ham:pi+stag}).
We have shown that the description of the spontaneously dimerized (SD) phase, including
its boundaries with the band insulator (BI) phases, can be well approximated by a sum
of two effective double-frequency sine-Gordon (DSG) models subject to self-consistency conditions
that couple the charge and spin sectors.
We have argued that, in the region $m_+ > m_-$ the quantum Ising transition to the CDW-like BI
phase is described
by the "charge" DSG model, 
whereas the criticality in "spin" DSG model is avoided. At $m_+ < m_-$
the situation is just the opposite. Using the well-studied critical properties of the DSG 
model \cite{DM},\cite{FGN1},\cite{FGN2} 
we found the critical exponent 1/8 characterizing the power-law
decay of the dimerization order parameter near the Ising criticality. We have also shown that
the staggered compressibility of the system near the SD-CDW transition, as well as the
staggered spin susceptibility at the SD-SDW transition display logarithmic singularities.

We have also discussed the topological excitations of the model
and derived the fractional values the charge $Q$ and $z$-projection of the spin $S^z$
they carry. We have traced the evolution of the topological kinks when moving from
the SD phase to one of the BI phases.

The results obtained in this paper also apply to the Peierls-Hubbard model
which includes a Coulomb onsite repulsion between the electron under the condition
that the latter is small compared to the attraction induced by the electron-phonon coupling.
A detailed description of the Peierls-Hubbard chain  perturbed by a spin-dependent staggered potential
is the subject of future studies.

\acknowledgements

We are grateful to M. Dalmonte,  G. Mussardo
and A.M. Tsvelik for their interest in this work and helpful comments.

\appendix

\section{Some details of Abelian bosonization}\label{bosrules}

Here we provide a brief account of the Abelian bosonization rules used in the main text
of the paper. For more details we refer to Ref. [\onlinecite{gnt}].

The model of free massless fermions with spin-1/2 
\be
{\cal H}^{(0)}_f (x) = - \ri v \sum_{\s=\up,\down}\psi^{\dagger}_{\s} (x)\tau_3 \p_x \psi_{\s} (x),
~~\psi_{\s} (x) = \left( 
\begin{array}{clcr}
R_{\s} (x)\\
L_{\s} (x)
\end{array}
\right)
\nn
\ee
is equivalent to a theory of two massless bosons:
\bea
 {\cal H}^{(0)}_b &=& \frac{v}{2} \sum_{\s = \up,\down}\left[ 
\Pi^2 _{\s} (x) + \left( \p_x \Phi_{\s} (x) \right)^2  \right] \nn\\
&=& \frac{v}{2} \sum_{a=c,s} \left[ 
\Pi^2 _{a} (x) + \left( \p_x \Phi_{a} (x) \right)^2 \right]
\label{FB}
\eea
where the charge  and spin  scalar fields ($\Phi_c, \Phi_s$) and their conjugate momenta
($\Pi_c, \Pi_s$) are defined as
\[
\Phi_{c,s} = \frac{\Phi_{\up} \pm \Phi_{\down}}{\sqrt{2}},
~~~\Pi_{c,s} = \frac{\Pi_{\up} \pm \Pi_{\down}}{\sqrt{2}}
\]
The smooth parts of the fermionic charge and spin densities are given by
\be
\rho(x) = \sqrt{\frac{2}{\pi}}\p_x \Phi_c (x), ~~~~~
S^z (x) = \frac{1}{\sqrt{2\pi}}\p_x \Phi_{c} \label{charge-spin}
\ee
Using the correspondence
\[
R^{\dagger}_{\s} L_{\s} \to - ({\ri}/{2\pi \alpha}) e^{- i \sqrt{4\pi} \Phi_{\s}}, ~~~
L^{\dagger}_{\s} R_{\s} \to  ({\ri}/{2\pi \alpha}) e^{i \sqrt{4\pi} \Phi_{\s}}
\]
one finds bosonized expressions for the
dimerization operator
\bea
D(x) &=& \sum_{\s}\bar{\psi}_{\s} \psi_{\s} =  \sum_{\s }\psi^{\dagger}_{\s} (x)\htau_2 
\psi_{\s}(x)\nn\\
&=& - \frac{1}{\pi \alpha} \sum_{\s} \cos \sqrt{4\pi} \Phi_{\s}\nn\\
&=&  - \left(\frac{2}{\pi \alpha}\right) \cos \sqrt{2\pi} \Phi_c (x) \cos \sqrt{2\pi} \Phi_s (x)
\label{dim-boson}
\eea
and the
staggered parts of charge and spin densities
\bea
 &&a^{-1}_0 (-1)^n  \sum_{\s} c^{\dagger}_{n\s} c_{n\s}  \to
\sum_{\s} \psi^{\dagger}_{\s} (x)\htau_1 
\psi_{\s} \nn\\
&=& - \frac{1}{\pi\alpha} \sum_{\s} \sin \sqrt{4\pi} \Phi_{\s}\nn\\
&=& - \left( \frac{2}{\pi \alpha}  \right)
\sin \sqrt{2\pi} \Phi_c (x) \cos \sqrt{2\pi} \Phi_s (x) \label{cdw-bos}\\
&& a^{-1}_0(-1)^n  \sum_{\s} \s c^{\dagger}_{n\s} c_{n\s}  \to \sum_{\s}\s \psi^{\dagger}_{s} (x)\htau_1  \psi_{\s}\nn\\
&=& - \frac{1}{\pi\alpha} \sum_{\s} \s \sin \sqrt{4\pi} \Phi_{\s}\nn\\ 
&=& - \left( \frac{2}{\pi \alpha}  \right)
\cos \sqrt{2\pi} \Phi_c (x) \sin \sqrt{2\pi} \Phi_s (x) \label{sdw-bos}
\eea
where $\alpha$ in a short-distance cutoff of the bosonic theory.

Squaring the dimerization operator (\ref{dim-boson}) yields the interaction term in 
the GN model (\ref{GN-L}). Here one makes
point splitting~
$
\left({\rm Tr}~\bar{\psi} \psi\right)^2 
\to  D(x) D(x+\alpha)
$
and uses short-distance Operator Product Expansion (OPE) \cite{cft,cft1}.
Employing the well-known OPE for the vertex operators
of a free Gaussian field $\Phi$
\bea
&&:e^{i\beta \Phi (z,\bz)}: :e^{i\beta' \Phi (0,0)}: \nn\\
&&~=~ \left(  \frac{|z|}{\alpha} \right)^{\frac{\beta\beta'}{2\pi}}
:e^{i[\beta \Phi(z,\bz) + \beta' \Phi(0,0)]}:
\label{basic-ope}
\eea
(here $z=v\tau + \ri x$, $\bar{z} = v\tau - \ri x$ and the symbol $::$ stands for
normal ordering) and keeping only axially symmetric
(in 1+1 dimensions -- Lorentz invariant) terms, 
one derives the following expansion 
\bea
&&\cos \beta \Phi (z,\bz) \cos  \beta \Phi (w,\bw)
= \frac{1}{2}\left(\frac{\alpha}{|z-w|}\right)^{\beta^2/2\pi} \nn\\
&&-\frac{1}{2}\beta^2 \alpha^{\beta^2/2\pi} |z-w|^{2-\beta^2/2\pi}
\p \varphi(w) \bp \bar{\varphi} (\bw)\nn\\
&&+ \frac{1}{2}\left( \frac{|z-w|}{\alpha}\right)^{\beta^2 /2\pi} \cos 2\beta \Phi(w,\bw) + \cdots~~~
\label{cft-cos-ope-gen}
\eea
where $\varphi(w)$ and $\bar{\varphi}(\bw)$ are holomorphic and anti-holomorphic
components of the scalr field $\Phi(w,\bw) = \varphi(w) + \bar{\varphi}({\bw})$,
and the dots stand for terms proportional to higher power of the distance $|z-w|$.
Setting $\beta^2 = 2\pi$ and $|z-w| = \alpha$ one obtains
\bea
\cos^2 \sqrt{2\pi} (x) &\to& \cos \sqrt{2\pi} \Phi(x) \cos \sqrt{2\pi} \Phi(x + \alpha) \nn\\
&=& \frac{1}{2} - (\pi\alpha)^2 \Big[
\frac{1}{8\pi} \left( (\p_x \Phi)^2 - \Pi^2  \right)\nn\\ 
&&- \frac{1}{(2\pi\alpha)^2}
\cos \sqrt{8\pi} \Phi
\Big] + \cdots
\label{co2-expan}
\eea
In fact, the expression in square brackets in (\ref{co2-expan}) is the Abelian version 
of the scalar product ${\bf J}_R \cdot {\bf J}_L $, where ${\bf J}_{R,L}$ are chiral vector currents of the critical SU(2)$_1$ 
Wess-Zumino-Novikov-Witten (WZNW) model (see e.g. [\onlinecite{gnt}])
\be
{\cal H}_{\rm WZNW} (x) = \frac{2\pi v}{3}
\left[
:{\bf J}^2 _R (x) : + :{\bf J}^2 _L (x):
\right]\label{crit-wzw}
\ee

\section{Minima of the potential $U(\Phi_c, \Phi_s)$}
\label{minima}

It is convenient to use the ($\Phi_{\up}, \Phi_{\down}$)-representation
and rewrite the potential (\ref{ham:pi+stag}) 
in the following dimensionless form

\bea
&& U(\Phi_{\up}, \Phi_{\down}) =  \left( \frac{g_0}{\pi \alpha} \right)^2
{\cal U} (x,y), \nn\\
&& {\fU}(x,y) = - \cx \cy - a \sx - b \sy \label{POT}
\eea
with
\bea
x = \sqrt{4\pi} \Phi_{\up}, ~~~
y = \sqrt{4\pi} \Phi_{\down}, \nn\\
a = m_{\up} \pi \alpha/g^2 _0, ~~~~~ b = m_{\down} \pi \alpha/g^2 _0
\eea
Here $a>0$ whereas the sign of $b$ is arbitrary. The potential ${\fU} (x,y)$
is periodic in $x$ and $y$ with a period $2\pi$, it has also a reflection
symmetry
\[
{\fU}(x,y) = {\fU} (\pi - x, \pi - y),
\]
as well as the $b \to - b$ symmetry:
\be
{\fU}(x,y; a, - b) = {\fU} (x, -y; a,b)
\label{b-b}
\ee

Classification of the extrema of the function ${\fU}(x,y)$ is standard. Introduce
the first and second partial
derivatives of the two-dimensional potential (\ref{POT}) :
\bea
&& {\fU}_x = \sx \cy - a \cx, ~~~~~ {\fU}_y = \cx \sy - b\cy \nn\\
&& {\fU}_{xx} = \cx \cy + a \sx, ~~~~~ 
{\fU}_{yy} = \cx \cy + b \sy, \nn\\
&&~~~~~~~~~~~~~~~~~~~~~{\fU}_{xy} = - \sx \sy \label{eq345}
\eea
Let ${\bf r}_0 = (x_0, y_0)$ be a solution of
the equations ${\fU}_x = 0$, ~${\fU}_y = 0$.
To identify this point introduce the quantity 
$
D = {\fU}_{xx} {\fU}_{yy} - {\fU}^2 _{xy}.
$
Then at $D>0$ ${\bf r}_0$ is a local minimum if  ${\fU}_{xx} > 0$ or a maximum
if ${\fU}_{xx} < 0$. At $D<0$ ${\bf r}_0$ is a saddle point.

At any $a$ and $b$ the equations ${\fU}_x = 0$ and ${\fU}_y = 0$
have solutions
$\left( {\pi}/{2}, \pm {\pi}/{2}  \right)$ and
$\left( -{\pi}/{2}, \pm {\pi}/{2}  \right)$. Using the above classification
rules, one finds that
the solution $(\pi/2, \pi/2)$ is a minimum
at $b>0$ if $ab > 1$ and a saddle point at $ab<1$ for arbitrary sign of $b$.
By the $b\to - b$ symmetry, the solution  $(\pi/2, - \pi/2)$ 
is a minimum at $b<0$ if $a|b| > 1$ and a saddle point at $a|b|<1$ for any sign of $b$.
A similar analysis shows that the solutions $(-\pi/2, \pm \pi/2)$ are either maxima of the potential
or saddle points. For this reason these solutions can be discarded.

Having found the local minima of the potential at $a|b|>1$, we now turn
to the case $a|b| < 1$. At $|x|, |y| \neq \pi/2$ the minima can be found from the equations
\be
\tan x = \frac{a}{\cy}, ~~~~~\tan y = \frac{b}{\cx}
\label{equas}
\ee
By the symmetry properties of the potential it is sufficient to consider the solutions in the first quadrant
$
0 < x < \pi, ~0 < y < \pi.
$
A simple analysis shows that at $b>0$ the minima of ${\fU}(x,y)$ 
appear in pairs, $(x_0, y_0)$ and $(\pi - x_0, \pi - y_0)$, with $0<x_0, y_0 < \pi/2$,
symmetrically located with respect to the point $(\pi/2, \pi/2)$:
\bea
x_0 = \arccos \sqrt{\frac{1 - a^2 b^2}{1 + a^2}},
~~y_0 = \arccos \sqrt{\frac{1 - a^2 b^2}{1 + b^2}}~~~
\label{roots}
\eea
The other set of minima is obtained from the previous one using the symmetry
(\ref{b-b}): $b\to -b$, $y \to - y$, $x \to x$.

At $a|b| \to 1$ $\varphi_+ \to \pi/2$, $\varphi_- \to 0$, and
the pairs of minima $(\vvphi^{(1)}, \vvphi^{(1)})$ merge at the points 
$(\frac{\pi}{2}+ \pi n_c, \pi n_s)$ at $m_+ > m_-$
and $(\pi n_c, \frac{\pi}{2}+ \pi n_s)$ at $m_+ < m_-$ [see Fig.\ref{motion}
and Eqs. (\ref{vac-a}), (\ref{vac-b})].

\section{Correction to effective action for spin-$\down$ electrons}\label{inter}

The part of the Euclidian action that accounts for interaction between the 
electrons with opposite spins is
\[
\Delta S = - \frac{g^2 _0}{v_F} \int \rd^2 {\bf r} ~
\left(   \psi^{\dagger}_{\up} \tau_2 \psi_{\up} \right)
\left(   \psi^{\dagger}_{\down} \tau_2  \psi_{\down} \right)
\]
The mass bilinears $\psi^{\dagger}_{\up} \tau_1 \psi_{\up}$
and $\psi^{\dagger}_{\up} \tau_2 \psi_{\up}$ have different parity properties. As a result,
in the lowest (i.e. zero) order
$
\la \psi^{\dagger}_{\up} \tau_2 \psi_{\up}\ra_{\up} = 0.
$
Therefore a nonzero correction to the spin-$\down$ part of the action 
appears 
in the second order in $g^2_0$:
\bea
&& \Delta S^{({\rm eff})}_{\down} = - \frac{1}{2} \la \left( \Delta S \right)^2 \ra_{\up}
= - \frac{g^4 _0}{2 v^2 _F} 
\int \rd^2 {\bf r}_1  \int \rd^2 {\bf r}_2 
\nn\\
&& 
\times \Big\la \left(   :\psi^{\dagger}_{\up} \tau_2 \psi_{\up}: \right)_{{\bf r}_1}
\left(   :\psi^{\dagger}_{\up} \tau_2 \psi_{\up}: \right)_{{\bf r}_2}
\Big\ra_{\up}\nn\\
&& \times
~\left(  : \psi^{\dagger}_{\down} \tau_2 \psi_{\down}: \right)_{{\bf r}_1}
\left(   :\psi^{\dagger}_{\down} \tau_2 \psi_{\down} :\right)_{{\bf r}_2}
\label{correction-eff}
\eea
The average in (\ref{correction-eff}) can be estimated using 
the 2$\times$2 Green's
function  matrix of the massive spin-$\up$ Dirac fermion:
\bea
&& \hat{\cal G}_{\up}(k,\vare) = \left( \ri \vare - \hat{\cal H}_{k\up}  \right)^{-1}
=
 - \frac{ \ri \vare +  kv_F \hat{\tau}_3 + m_{\up} \hat{\tau}_1 }{\vare^2 + k^2 v^2 _F 
 + m^2 _{\up}},~~~
\label{GF-matrix}
\eea
Choosing $\tau > 0, ~m_{\up} >0$ we obtain:
\bea
&& \hat{G}_{\up} ({\bf r}) =  \int \frac{\rd k}{2\pi} \int \frac{\rd \vare}{2\pi}
e^{ikx - i \vare \tau} \hat{\cal G}_{\up}(k,\vare)
\nn\\
&&= - \left(\frac{m_{\up}}{2\pi v_F} \right)\Big[
\frac{v_F \tau + \ri x \hat{\tau}_3}{r} K _1 (m_{\up}r/v_F)\nn\\
&&~~~~~~~~~~~~~~~~~~~~~~~~~~~+~ \hat{\tau}_1 K_0 (m_{\up}r/v_F)
\Big]\label{G-fin}
\eea
where $K_0 (z)$ is the Macdonald function, $K_1 (z) = - K' _0 (z)$ and
$r = \sqrt{x^2 + v^2 _F \tau^2} $.
The average in (\ref{correction-eff}) 
\bea
&&X_{\up}({\bf r}) \equiv \la   :\psi^{\dagger}_{\up} ({\bf r}) \tau_2 \psi_{\up} ({\bf r}): 
:\psi^{\dagger}_{\up}({\bf 0}) \tau_2 \psi_{\up}({\bf 0}) : \ra_{\up} \nn\\
&&= - {\rm Tr}~\left[ \hat{\tau}_2 \hat{G}_{\up} ({\bf r}) \hat{\tau}_2  \hat{G}_{\up} (-{\bf r})  \right], 
\nn\\
&& = \frac{\Delta^2_{\up}}{2\pi^2 v^2 _F}
\left[ K^2 _0 (m_{\up}r/v_F) + K^2 _1 (m_{\up}r/v_F)\right]
\label{X}
\eea
represents a 
polarization loop, that is 
the uniform static susceptibility
of the massive spin-$\up$ fermions with respect to dimerization of the system.
Since the mass term in the Hamiltonian ${\cal H}_{\up}$ has a $\hat{\tau}_1$
structure whereas the dimerization operator has a $\hat{\tau}_2$
structure, 
the integral
\be
\chi_{{\rm D}{\up}} = \int \rd^2 {\bf r} ~X_{\up}({\bf r})
\label{chi-D}
\ee
represents the static uniform limit of 
the "transverse" susceptibility.

Eq.(\ref{correction-eff}) can be compactly rewritten as follows:
\bea
&& \Delta S^{({\rm eff})}_{\down} =\nn\\
&& -  \frac{g^4 _0}{2 v^2 _F} 
\int \rd^2 {\bf r}_1  \int \rd^2 {\bf r}_2~
X_{\up}({\bf r}_1-{\bf r}_2) ~D_{\down} ({\bf r}_1) D_{\down}({\bf r}_2)~~~
\label{correction}
\eea
where 
\be
D_{\down}({\bf r}) ~=~ :\psi^{\dagger}_{\down} ({\bf r}) \hat{\tau}_2 
\psi_{\down}({\bf r}):
\label{down-dim}
\ee
is the dimerization operator for the spin-$\down$ fermions.
As follows from (\ref{G-fin}), at distances larger than the correlation length
$\xi_{\up} \sim v_F/m_{\up}$,
the integral kernel $X_{\up}({\bf r}_1-{\bf r}_2)$ decays exponentially. 
Having in mind that the characteristic correlation length of the spin-$\down$ is much larger,
$\xi_{\down} \sim v_F /m_{\down}\gg \xi_{\up}$, it is legitimate to
treat the product of normal ordered dimerization operators, 
$D_{\down} ({\bf r}_1) D_{\down}({\bf r}_2)$,
by means of short-distance {Operator Product Expansion} (OPE) \cite{cft}
(see Appendix \ref{bosrules}). To accomplish this
procedure, introduce new coordinates:
$
{\bf x} = ({\bf r}_1 + {\bf r}_2)/2$, ${\bf r} = {\bf r}_1 - {\bf r}_2$.
Using the bosonic 
representation of the dimerization  field,
$
D_{\down} = - {(\pi\alpha)^{-1}} \cos \sqrt{4\pi} \Phi_{\down},
$
and the Euclidian version of the OPE (\ref{cft-cos-ope-gen}) for $\beta^2 = 4\pi$, we arrive at the
fusion rule of two dimerization operators:
\bea
 D_{\down} ({\bf r}_1) D_{\down}({\bf r}_2) &=& \frac{1}{(\pi\alpha)^2}
:\cos \sqrt{4\pi} \Phi_{\down} ({\bf r}_1): :\cos \sqrt{4\pi} \Phi_{\down} ({\bf r}_2): 
\nn\\
&=& {\rm const} - \frac{1}{2\pi}
 \left( \vnab \Phi_{\down}({\bf x}) \right)^2 \nn\\
&+& \frac{1}{2(\pi\alpha)^2} \left(\frac{r}{\alpha}\right)^2\cos \sqrt{16\pi} \Phi_{\down}({\bf x})
\label{DD-ope}
\eea
When (\ref{DD-ope}) is substituted into (\ref{correction}), one obtains:
\begin{widetext}
\bea
\Delta S^{({\rm eff})}_{\down} &=&  \frac{g^4 _0}{2 v^2 _F} 
\int \rd^2 {\bf r}~ X_{\up}({\bf r}) \int \rd^2 {\bf x}~
\left[
\frac{1}{2\pi}
 \left( \vnab \Phi_{\down}({\bf x}) \right)^2 
- \frac{1}{2(\pi\alpha)^2} \left(\frac{r}{\alpha}\right)^2\cos \sqrt{16\pi} \Phi_{\down}({\bf x})
\right] 
 \nn\\
&=& \int \rd {\bf x}~\left[ \frac{1}{2}A  \left( \vnab \Phi_{\down} ({\bf x})  \right)^2
- \frac{B}{\alpha^2}  \cos \sqrt{16 \pi} \Phi ({\bf x}) \right]
\label{correction1}
\eea
\end{widetext}
where
\bea
A &=&  \frac{g^4 _0}{2\pi v_F} \chi_{{\rm D}{\up}}
= \frac{1}{2}\left( \frac{g^2 _0}{\pi v_F}  \right)^2 \ln \frac{\Lambda}{m_{\up}},
\label{A}\\
B &=& C_1 \left( \frac{g^2 _0}{v_F} \right)^2 \left( \frac{\Lambda}{m_{\up}}  \right)^2,
\label{AB}
\eea
$C_1$ being a positive numerical constant.

Let us comment on the structure of the parameter $A$ in (\ref{A}). In the massless limit 
($m_{\up} = 0$)
the dimerization susceptibility $\chi_{D\up}$ of the spin-$\up$ fermions represents a 
particle-hole loop with the frequency-momentum transfer 
$\omega=0$, $q = \pi/a_0$
and is
logarithmically divergent. For fermions with a small nonzero mass $m_{\up} \ll \Lambda$
the infrared logarithmic divergency of $\chi_{D\up}$
is cut off  by the Dirac mass:
$
\chi_{D\up} \simeq (\pi v_F)^{-1} \ln ({\Lambda}/{|m|}). 
$
The logarithmic integration in (\ref{chi-D}) goes over the 
short-distance region $a_0 \ll |x|, v_F |\tau| \ll \xi$
($\xi \sim v_F / m$) where perturbative single-particle renormalizations take place
 (in this region
$X(r) \sim K^2 _1 (mr/v_F) \sim 1/r^2$).


\newpage
\begin{thebibliography}{99}

\bibitem{nagaosa} N. Nagaosa and J. Takimoto, J. Phys. Soc. Jpn.
{\bf 55}, 2735 (1986); N. Nagaosa, {\it ibid.} {\bf 55},2754
(1986); {\bf 55}, 3488 (1986).
\bibitem{egami}
T. Egami, S. Ishihara, and M. Tachiki, Science {\bf 261}, 1307 (1993).
\bibitem{FGN1} M. Fabrizio, A.O. Gogolin, A.A. Nersesyan, Phys. Rev. Lett. {\bf 83}, 2014 (1999).
\bibitem{FGN2} M. Fabrizio, A.O. Gogolin, A.A. Nersesyan,
Nucl.Phys.B {\bf 580} [FS] 647 (2000);
A.A. Nersesyan, ÓIsing-model description of Quantim Critical Points in 1D Electron and Spin SystemsÓ,in NATO ASI/EC Summer School: New Theoretical Approaches to Strongly Correlated Systems, 
pp. 93Ð120, Kluwer Academic Publishers, 2001.
\bibitem{manmana} S.R. Manmana, V. Meden, R.M. Noack, and K. Sch\"{o}nhammer, Phys. Rev.
B {\bf 70}, 155115 (2004).
\bibitem{zhang} Y. Z. Zhang, C. Q. Wu, and H. Q. Lin, Phys. Rev. B {\bf 72}, 125126  (2005). 
\bibitem{dyonis} L. Tincani, R.M. Noak, and D. Baeriswyl, Phys. Rev. B {\bf 79}, 165109 (2009).
\bibitem{tarruell} L. Tarruell, D. Greif, T. Uehlinger, G. Jotzu, and  T. Esslinger,
Nature {\bf 483}, 302 (2012).
\bibitem{exp1} T. Uehlinger, G. Jotzu, M. Messer, D. Greif, 
W. Hofstetter, U. Bissbort, and T. Esslinger, Phys. Rev. Lett.
{\bf 111}, 
185307 (2013).
\bibitem{messer} M. Messer, R. Desbuquois, T. Uehlinger,
G. Jotzu, S. Huber, D. Greif,
and T. Esslinger, Phys. Rev. Lett. {\bf 115}, 115303 (2015).
\bibitem{orignac}
K. Loida, J.-S. Bernier,  R. Citro, E. Orignac, and C. Kollath,
Phys. Rev. Lett. {\bf 119}, 230403 (2017).
\bibitem{nakamura} M. Nakamura, J. Phys. Soc. Jpn. {\bf 68}, 3123 (1999);
Phys. Rev B {\bf 61}, 16377 (2000).
\bibitem{furu} M. Tsuchiizu and A. Furusaki, Phys. Rev. Lett. {\bf 88}, 056402 (2002);
PR B{\bf 69}, 035103 (2004).
\bibitem{dender} D. C. Dender,  P. R. Hammar,  D. H. Reich,  and 
C. Broholm, and G. Aeppli, Phys. Rev. Lett. {\bf 79}, 1750 (1997).
\bibitem{aff-oshi} M. Oshikawa and I.Affleck, Phys. Rev. Lett. {\bf 79}, 2883 (1997).
\bibitem{ess-tsv} F.H.L. Essler and A.M. Tsvelik, Phys. Rev. B{\bf 57}, 10 592 (1998).
\bibitem{zheludev}
A. Zheludev et al., Phys. Rev. Lett. {\bf 80}, 3630  1998;
20S. Maslov and A. Zheludev, Phys. Rev. Lett. 80, 5786  1998 ;
Phys. Rev. B 57, 68  1998 .
\bibitem{Peierls} R.E. Peierls, {\sl Quantum Theory of Solids}, Oxford University
Press, London, 1955.
\bibitem{SSH} W.P. Su, J.R. Schrieffer and A.J. Heeger, Phys. Rev. Lett. {\bf 42},
1698 (1979); PR B {\bf 22}, 2099 (1980).
\bibitem{ssh2} A.J. Heeger, S. Kivelson, J.R. Schrieffer, and W.P. Su, Rev. Mod. Phys. {\bf 60}, 781 (1988).
\bibitem{braz1} S.A. Brazovskii, JETP Lett. {\bf 28}, 606 (1978);
Sov.Phys. JETP {\bf 51}, 677 (1980).
\bibitem{ns} A. Niemi and G.W. Semenoff, Phys. Rept. {\bf 135}, 100 (1986).
\bibitem{bk} S.A. Brazovskii and N.N. Kirova, JETP Lett. {\bf 33}, 4 (1981).
\bibitem{bkm} S.A. Brazovskii, N.N. Kirova and S.I. Matveenko, Sov. Phys. JETP {\bf 59},
424 (1984).
\bibitem{shen} S.-Q. Shen, {\sl Topological Insulators}, Springer, 2012
\bibitem{essler-2006} H. Benthien, F.H.L. Essler, and A. Grage, Phys. Rev. B{\bf 73}, 085105 (2006).
\bibitem{tll} H. Takayama, Y.R. Lin-Liu, and K. Maki, Phys.Rev. {\bf 21}, 2388 (1980).
\bibitem{BD} S.A. Brazovskii and I.E. Dzyaloshinskii, Sov. Phys. JETP {\bf 44}, 1233 (1976).
\bibitem{fh1} E. Fradkin and J.E. Hirsch, Phys. Rev. B {\bf 27}, 1680 (1983).
\bibitem{AGD} A.A, Abrikosov, L.P. Gor'kov, and I.E. Dzyaloshinskii, {\sl Methods of
Quantum Field Theory in Statistical Physics}, ed. R.A. Silverman, Dover, New York
(1963).
\bibitem{LRA} P.A. Lee, T.M. Rice and P.W. Anderson, Phys. Rev. Lett. {\bf 31}, 462
(1973); Solid State. Commun. {\bf 14}, 703 (1974).
1233 (1976).
\bibitem{cardy} J. Cardy, {\sl Scaling and Renormalization in Statistical Physics},
Cambridge University Press, 1996.
\bibitem{jr} R. Jackiw and C. Rebby, Phis. Rev. D {\bf 13},3398 (1976).
\bibitem{gw} J. Goldstone and F. Wilczek, Phis. Rev. Lett. {\bf 47}, 986 (1981).
\bibitem{GN} D. Gross and A. Neveu, Phys. Rev. D {\bf 10}, 3235 (1974).
\bibitem{witten1} E. Witten, Nucl.Phys. B {\bf 142}, 285 (1978).
\bibitem{shankarwitten} R. Shankar and E. Witten, Nucl.Phys. B {\bf 141}, 349 (1978).
\bibitem{DHN} R. Dashen, B. Hasslacher, and A.Neveu, Phys. Rev. D{\bf 12}, 2443 (1975).
\bibitem{zamo2} A.B. Zamolodchikov and Al.B. Zamolodchikov, Ann.Phys. {\bf 120}, 253 (1979).
\bibitem{gnt}  A.O. Gogolin, A.A. Nersesyan and A.M. Tsvelik, 
{\sl Bosonization and Strongly Correlated Systems}, Cambridge University Press, 1998.
\bibitem{k} J.Kogut, Rev. Mod. Phys. {\bf 51}, 659 (1979)
\bibitem{LZ1} S. Lukyanov and A. Zamolodchikov,  Nucl. Phys. B {\bf 493}, 571(1997).
\bibitem{giam} T. Giamarchi, {\sl Quantum Physics in One Dimension}, Oxford University Press,
2003.
\bibitem{DM} G. Delfino, G. Mussardo, Nucl.Phys.B {\bf 516}, 675 (1998).
\bibitem{cft} G. Mussardo, {\sl Statistical Field Theory}, Oxford University Press, 2010.
\bibitem{cft1} P. di Francesco, P. Mathieu, D. S\'{e}n\'{e}chal, {\sl Conformal Field Theory}, Springer,
1997.
\bibitem{LZ2} S. Lukyanov and A. Zamolodchikov, Nucl.Phys. B {\bf 607}, 437 (2001).
\bibitem{coleman} S. Coleman, Phys. Rev. D {\bf 11}, 2088 (1975).
\bibitem{suzu} Y. Suzumura, Prog. Theor. Phys. {\bf 61}, 1 (1979).
\bibitem{fh2} J.E. Hirsch and E. Fradkin, Phys. Rev. B {\bf 27}, 4302 (1983).

\end{thebibliography}
\end{document}